\documentclass[reprint,aps,letterpaper,10pt,twoside=false]{revtex4}
\usepackage[left=3cm, right=3cm, bottom=2.8cm, top=3.1cm]{geometry}
\usepackage{graphicx}
\usepackage{dcolumn}
\usepackage{bm}
\usepackage{float}
\usepackage{amssymb}
\usepackage{booktabs}
\usepackage{caption}
\usepackage{subcaption}
\usepackage{subfloat}
\usepackage{braket}
\usepackage{amsmath} 
\usepackage{titlesec}
\usepackage{cleveref}

\titlespacing*{\subsubsection}{0pt}{1.5ex plus 0.2ex minus 0.1ex}{1.3ex}

\crefname{figure}{Fig.}{Figs.}

\begin{document}

\preprint{AAPM/123-QED}

\title{Spectro-temporally tailored Non-Gaussian Quantum Operations in Thin-Film Waveguides}

\author{
Peter Namdar$^{1}$\thanks{Corresponding author: \texttt{peter.namdar@lkb.upmc.fr}}, Patrick Folge$^2$, Carlos E. Lopetegui$^1$, Silia Babel$^2$,\\
Benjamin Brecht$^2$, Christine Silberhorn$^2$, Valentina Parigi$^1$\\[1.2ex]  
\textsuperscript{1}\textit{Laboratoire Kastler Brossel, Sorbonne Université, CNRS, ENS-Université PSL, Collège de France, 4 place Jussieu, F-75252 Paris, France}\\
\textsuperscript{2}\textit{Integrated Quantum Optics, Institute for Photonic Quantum Systems (PhoQS), Paderborn University, Warburgerstr. 100, Paderborn 33098, Germany}\\[0.8ex]
\textit{*Corresponding author: peter.namdar@lkb.upmc.fr}
}

\begin{abstract} 
Advancements in photonic platforms have enabled the precise control of light's spectral and temporal degrees of freedom, a capability crucial for the development of scalable quantum information systems. In this work, we address the challenge of implementing spectro-temporal mode-selective non-Gaussian quantum operations, specifically single-photon subtraction (SPS) and addition (SPA), in the telecom wavelength regime. Building on prior experimental demonstrations of mode-selective near-infrared SPS, we present the first design framework for achieving mode-selective SPA and SPS using thin-film lithium niobate nonlinear waveguide platforms.
We introduce an inverse-design optimization scheme by modeling the quantum-optical response via the Joint Spectral Amplitude (JSA) and Transfer Function (TF), in order to identify optimal waveguide and pump parameters that maximize mode selectivity and state purity. This approach is first tested on a metallic waveguide design. We then exploit the dispersion engineering capabilities of thin-film waveguides, which offer enhanced nonlinear interactions through tighter light confinement.
Our findings demonstrate that tailored nonlinear processes, particularly parametric down-conversion and frequency up-conversion, can support high-fidelity non-Gaussian operations essential for next-generation quantum photonic networks.
\end{abstract}

\maketitle

\section{Introduction}
Photonics platforms provide powerful and versatile tools for encoding and transmitting information in both classical and quantum protocols. To fully harness the potential of light multiplexing—encoding across the complete set of optical degrees of freedom—comprehensive spectro-temporal control is essential for advancing quantum applications \cite{Fabre_2020,Brecht2015}. In particular, multimode spectro-temporal encoding for quantum information processing necessitates precise manipulation of quantum operations in both the spectral and temporal domains. While significant progress has been made in generating spectrally and temporally multiplexed quantum states with Gaussian field statistics   \cite{Larsen19,Inoue23,Kashiwazaki23,Nehra22} operations exploiting non-Gaussian quantum field statistics remain costly and challenging to control \cite{Walschaers2021}.

Achieving spectro-temporal mode-selective operations at the single-photon level requires highly tailored nonlinear optical processes. Thin-film waveguides now serve as an effective platform for these processes \cite{Zhu_21, boes2023, Wang2025}. However, optimizing fabrication parameters to fine-tune the nonlinear response introduces a complex inverse-design problem.
In this work, we identify a set of optimal parameters for designing spectro-temporal mode-selective single-photon subtraction (SPS) and addition  (SPA) operations at telecom wavelengths via an optimization scheme.
Previous experiments have successfully demonstrated SPA and SPS using nonlinear media \cite{Zavatta_2004, Zavatta_2005, Zavatta_2007, Parigi_2007}, typically performed directly on the whole input fields to analyze their resulting (non-)classicality. While mode-selective SPS was experimentally realized in the near-infrared by Ra et al. in 2017 \cite{Ra17}, our work extends this concept into the telecom regime and additionally proposes the first implementation of mode-selective SPA in general.
While waveguides with micrometer-scale cross-sections have been implemented in various quantum optical experiments \cite{Obrien_2009, Qi_2020}, recent technological advancements have introduced an additional trend toward so-called thin-film waveguides. These structures confine light to significantly smaller cross-sections, often in the sub-micron scale, thereby greatly enhancing nonlinear interactions \cite{Zhu_21, Vazimali_2022}. 
Recent studies have shown that by carefully tailoring the waveguide geometry, the spectral properties of the interacting fields can be precisely engineered to achieve optimal phase-matching conditions, maximize mode overlap, and enhance conversion efficiency \cite{Koyaz_24, Arora_2024, Luo_18}. Advances in manufacturing and calibration techniques enable correction of fabrication imperfections, thereby making it feasible to realize theoretical configurations, such as those proposed in this study \cite{Xin25}. Additionally, significant progress has been made in developing low-loss thin-film waveguides in the telecom-wavelength regime, further enhancing the potential for high-performance integrated photonic systems \cite{Krasnokutska_18, Shams-Ansari2022,zhu_twenty-nine_2024, Kellner2025, sabatti2025}.\\
Starting from a fully analytical model that has been shown to be a sufficient guideline, we could later apply our schematics to thin-film lithium niobate as described in Sec. 2.4.b. For a quantum-optical description of the proposed experimental scenarios, we express the so-called Joint Spectral Amplitude (JSA) or Transfer Function (TF) of the nonlinear scheme, which is determined by the waveguide parameters and the incoming classical light field (see Sec. \ref{sec:general_description}). \\
To optimize these waveguide and pump-defining parameters, we introduce an algorithmic optimization approach designed to efficiently identify optimal configurations within the parameter space (see Methods). In this context, we establish the necessary conditions for valid solutions and analyze the resulting purity of the non-Gaussian states, providing both a description and evaluation.  We focus entirely on the well-studied regime of second-order nonlinear processes ($\chi^{(2)}$-processes), specifically parametric down-conversion (PDC) and frequency up-conversion. To model the mode-selective scenarios, we adopt established approaches, evolving the incoming states either through a beam splitter or via a nonlinear PDC Hamiltonian \cite{Averchenko14, Averchenko16, Roeland22}.\\ Given the established infrastructure of classical communication networks and the recent demonstration of multimode cluster states in the telecom regime, we focus our approach on applying non-Gaussian processes to light at a central wavelength of $\lambda_{center}=1560~nm$ \cite{Roman_Rodriguez_2021}. However, this scheme is wavelength-independent and can, in principle, be applied at any wavelength.\\ For clarification, in this work we always sketch a non-collinear configuration involving three interacting beams. In reality, these beams are assumed to be well-confined within a single spatial mode of a waveguide, where they can be distinguished by either wavelength or polarization. 

\section{Non-Gaussian processes on multimode quantum states}
\label{sec:general_description}

The two processes of SPA and SPS rely on nonlinear interactions in a $\chi^{(2)}$-medium, where an input quantum state interacts with a strong classical beam, facilitating controlled photon addition or subtraction in specific spectral-temporal modes. In the following, we provide a detailed description of each process, highlighting their theoretical foundations and experimental feasibility.

\begin{figure}[h!]
\centering
\begin{minipage}[c]{0.46\textwidth}
    \centering
    \includegraphics[width=\textwidth]{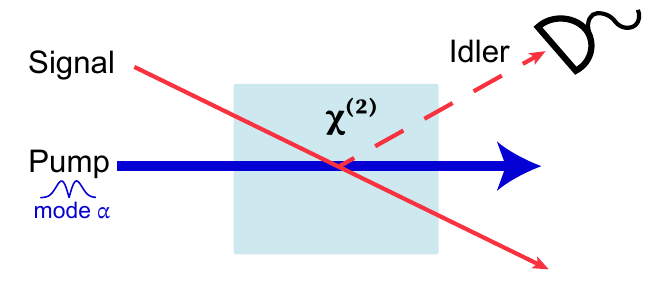}
    \caption*{(a) Single-photon addition (SPA)}
    \label{fig:scheme_SPA}
\end{minipage}
\hfill
\begin{minipage}[c]{0.51\textwidth}
    \centering
    \includegraphics[width=\textwidth]{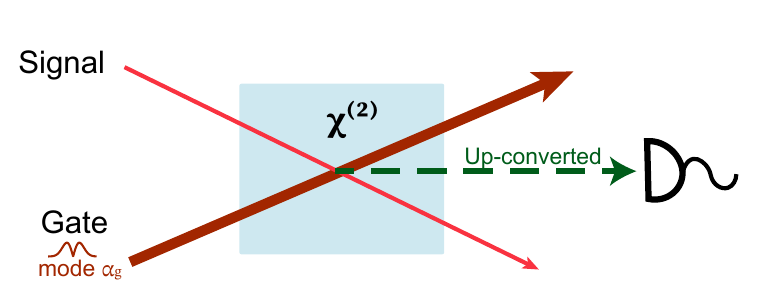}
    \caption*{(b) Single-photon subtraction (SPS)}
    \label{fig:scheme_SPS}
\end{minipage}

\vspace{1ex}

\caption{Schematic representations of the experimental setups discussed in this work. For clarity, the optical paths are illustrated in a non-collinear configuration, although in well-confined waveguides, they are collinear and can be separated by wavelength and/or polarization. The two processes are heralded by photon detection on one of the paths.}
\label{fig:experimental_setups}
\end{figure}

\subsection{Mode-Selective Single-Photon Addition}
\label{sec:SPA}
Fig. \ref{fig:experimental_setups}a illustrates a setup for performing single-photon addition on an incoming quantum state \(\hat{\rho}\). In this configuration, the two incident fields — the signal \(\hat{\rho}\) and a strong coherent pump — interact within a $\chi^{(2)}$-medium, generating a third beam via parametric down-conversion. The signal state is associated with quantum mode \(\ket{a}\), with the creation and annihilation operators \(\{\hat{a}^\dagger, \hat{a}\}\), while the up-converted beam is generated in the idler mode \(\ket{b}\), associated with $\{\hat{b}^\dagger, \hat{b}\}$ , which is initially prepared in the vacuum state \(\ket{0}\). The system’s nonlinear optical dynamics are governed by the Hamiltonian:
\[
\hat{H} = \int \int d\omega_s d\omega_i \, J(\omega_s, \omega_i) \hat{a}^\dagger(\omega_s) \hat{b}^\dagger(\omega_i) + \text{h.c.}
\]

where \(J(\omega_s, \omega_i)\) is the Joint Spectral Amplitude (JSA), describing the spectral correlations between the two emerging quantum fields.
\(J(\omega_s, \omega_i)\) can be expressed as the product of the pump beam's spectral shape \(\alpha(\omega_s + \omega_i)\) and the crystal's phase-matching function \(\phi(\omega_s, \omega_i)\), where \(\omega_s\) and \(\omega_i\) are the frequencies of the signal and idler photons \cite{Roman_Rodriguez_2021}. Since the JSA encodes the spectral correlations between these fields, analyzing its structure is crucial for understanding the mode composition of the generated quantum state and further the mode-selectivity of the process. A natural approach to achieve this is through Schmidt decomposition, which is mathematically equivalent to a singular-value decomposition (SVD) of the JSA. This decomposition provides eigenmodes and corresponding eigenvalues \(\lambda_l\), quantifying the contributions of specific mode pairs \(\{h_l(\omega_s), g_l(\omega_i)\}\) to the overall state. To diagonalize the JSA to obtain the eigenstates and their contrictuib, in 2000 C.K. Law extended the Schmidt decomposition framework to the spectro-temporal domain \cite{Law2000}.

\begin{equation}
    J(\omega_s, \omega_i) = \alpha(\omega_s + \omega_i) \phi(\omega_s, \omega_i) = \sum_l \sqrt{\lambda_l} h_l(\omega_s) g_l(\omega_i)
\end{equation}

The role of the phase-matching function \(\phi(\omega_s, \omega_i)\) is discussed in sec. 2.4, while the impact of the pump spectral shape \(\alpha(\omega_s + \omega_i)\) is detailed in sec. \ref{sec:purity}.

\subsection{Mode-selective Single-Photon Subtraction}

Fig. \ref{fig:experimental_setups}b illustrates a setup for implementing mode-selective single-photon subtraction from an incoming multimode state \(\hat{\rho}\). Similar to the SPA scheme from sec. 2.1, two incident light beams overlap in a $\chi^{(2)}$-medium, resulting in the generation of a third beam. In this configuration, the strong semi-classical beam is labelled as the gate field. The signal state is associated with mode \(\ket{a}\), while the up-converted field is associated with mode \(\ket{c}\), each characterized by the ladder operators \(\{\hat{a}^\dagger, \hat{a}\}\) and \(\{\hat{c}^\dagger, \hat{c}\}\), respectively. As photons are created in the up-converted beam, photons are simultaneously annihilated from the signal beam. This evolution can be described by a beam splitter Hamiltonian \cite{Walschaers2021}:

\begin{equation}
    \hat{H} = \int \int d\omega_s d\omega_{\text{up}} \, T(\omega_s, \omega_{\text{up}}) \hat{a}(\omega_s) \hat{c}^\dagger(\omega_{\text{up}}) + \text{h.c.},
\end{equation}

where \(T(\omega_s, \omega_{\text{up}})\) is the transfer function, analogous to the JSA in the single-photon addition scheme \cite{Averchenko14}. Therefore, \(T(\omega_s, \omega_{\text{up}})\) describes the frequency up-conversion mapping from signal frequencies to output frequencies:

\begin{equation}
\begin{split}
    T(\omega_s, \omega_{\text{up}}) &= \alpha_g(\omega_{\text{up}} - \omega_s) \phi(\omega_s, \omega_{\text{up}}) \\
    &= \sum_l \sqrt{\lambda_l} m_l(\omega_s) n_l(\omega_{\text{up}})
\end{split}
\end{equation}

This can be expressed as the product of the gate beam's spectral shape \(\alpha_g(\omega_{\text{up}} - \omega_s)\) and the phase-matching function \(\phi(\omega_s, \omega_{\text{up}})\), where \(\omega_s\) and \(\omega_{\text{up}}\) denote the frequencies of the signal and up-converted photons, respectively. This decomposition naturally reveals the mode structure of the process, where the eigenmodes \(\{m_l(\omega_s), n_l(\omega_{\text{up}})\}\) and their corresponding eigenvalues \(\lambda_l\) quantify the spectral contributions of individual eigenmode pairs. 

Notably, the transfer function and the coupling between the signal and up-converted fields are directly analogous to the quantum pulse gate, first investigated by C. Silberhorn's group \cite{Eckstein2011}. 

\subsection{Goal: Desired JSA and TF geometry}
\label{sec:schematic_goal}
\begin{figure}[h!]
  \centering
  \subfloat[HG$_0$\label{fig:JSA_schematic_HG0}]{%
    \includegraphics[width=0.3\linewidth]{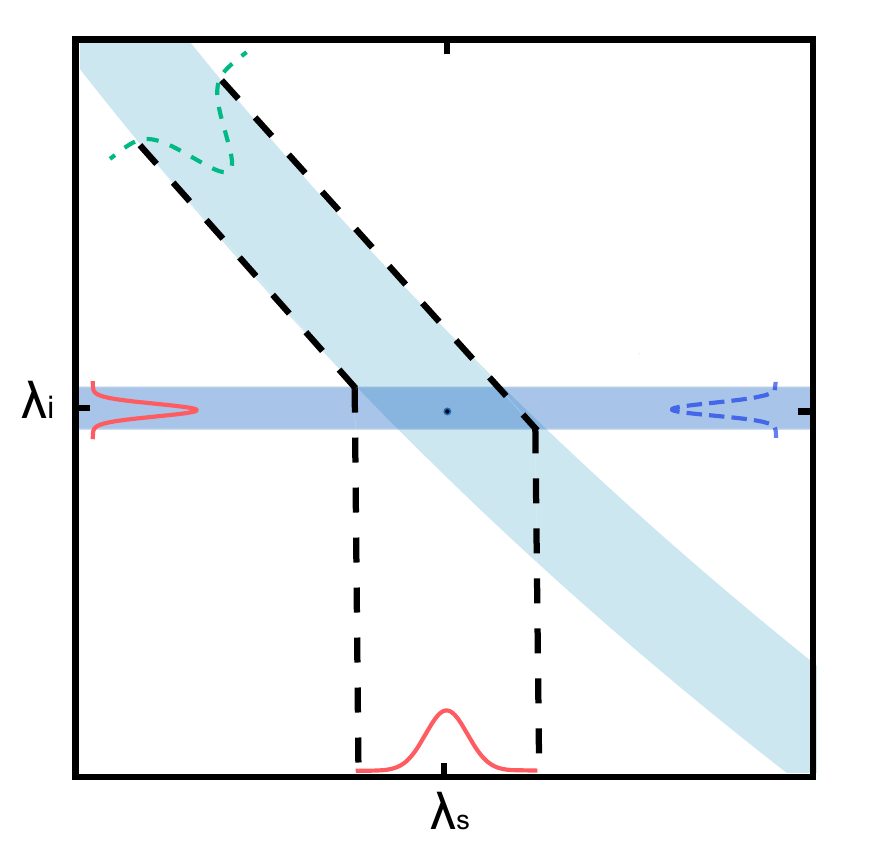}
  }%
  \hspace{2em}  
  \subfloat[HG$_1$\label{fig:JSA_schematic_HG1}]{%
    \includegraphics[width=0.3\linewidth]{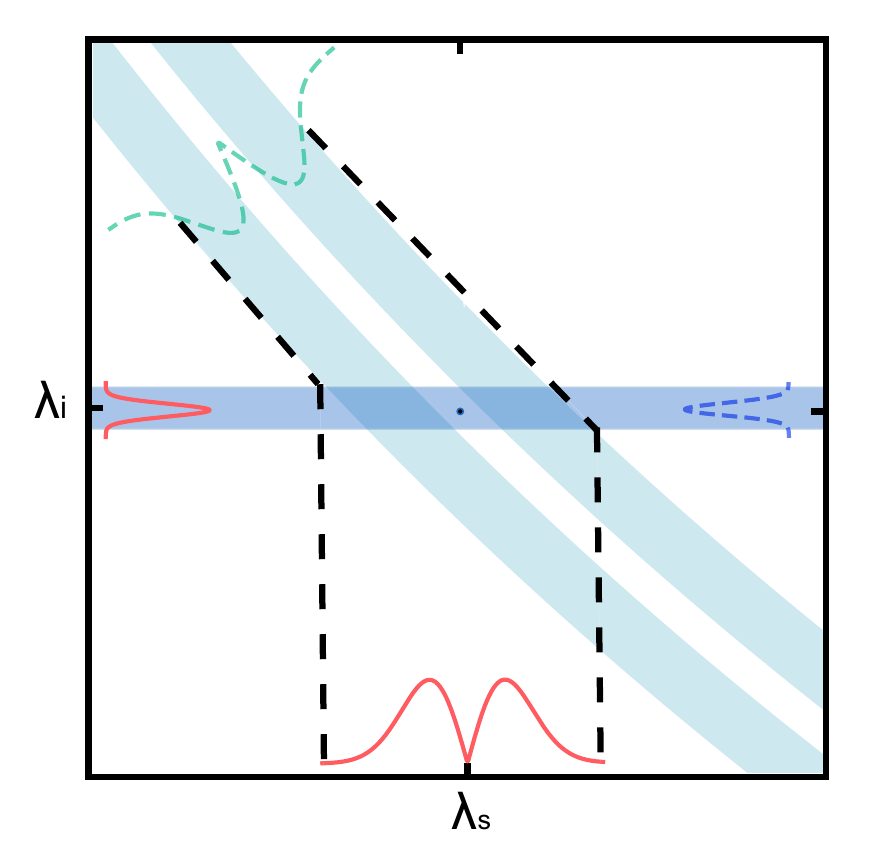}
  }
\caption{Schematic JSAs for mode-selective SPA. Here the incoming pump mode (dashed green line) gets transferred onto the signal axis (red signal mode) due to the intersection with the narrow, horizontal phase-matching function (blue area). There, the narrow and horizontal shape of the phase-matching function results (dashed blue line) in a narrow idler mode (red mode on the idler axis), suitable for heralding.}
  \label{fig:jsa_schematics}
\end{figure}

\begin{figure}[h!]
  \centering
  \subfloat[HG$_0$\label{fig:TF_schematic_HG0}]{%
    \includegraphics[width=0.3\textwidth]{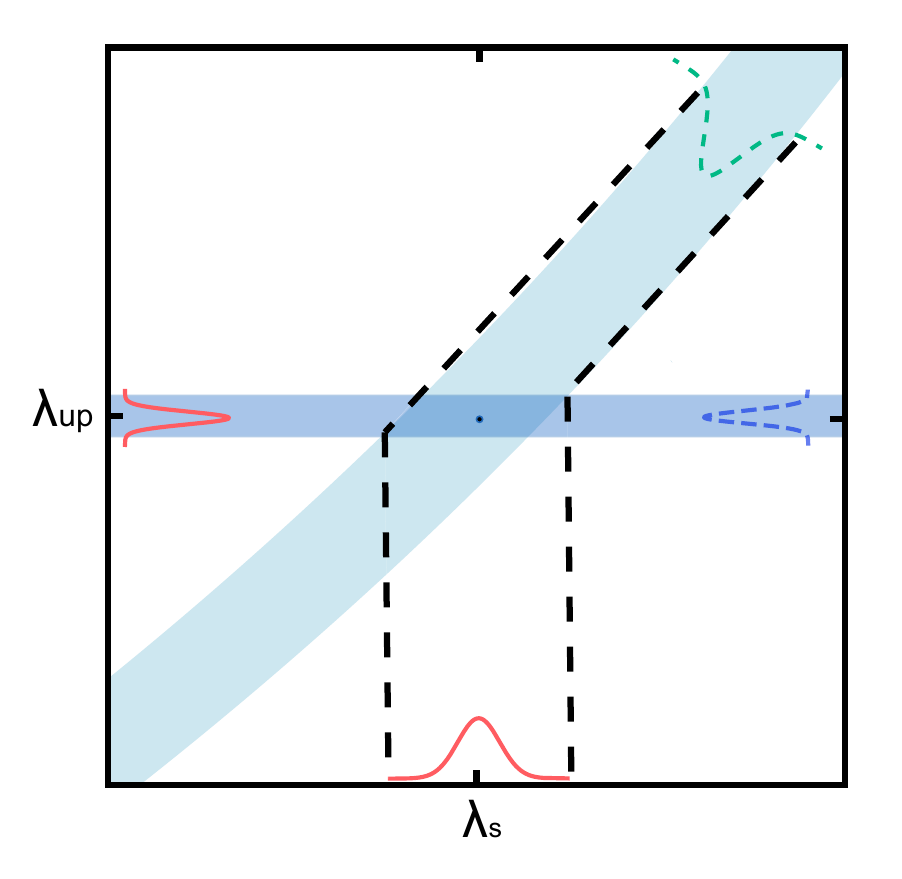}
  }
  \hspace{2em}
  \subfloat[HG$_1$\label{fig:TF_schematic_HG1}]{%
    \includegraphics[width=0.3\textwidth]{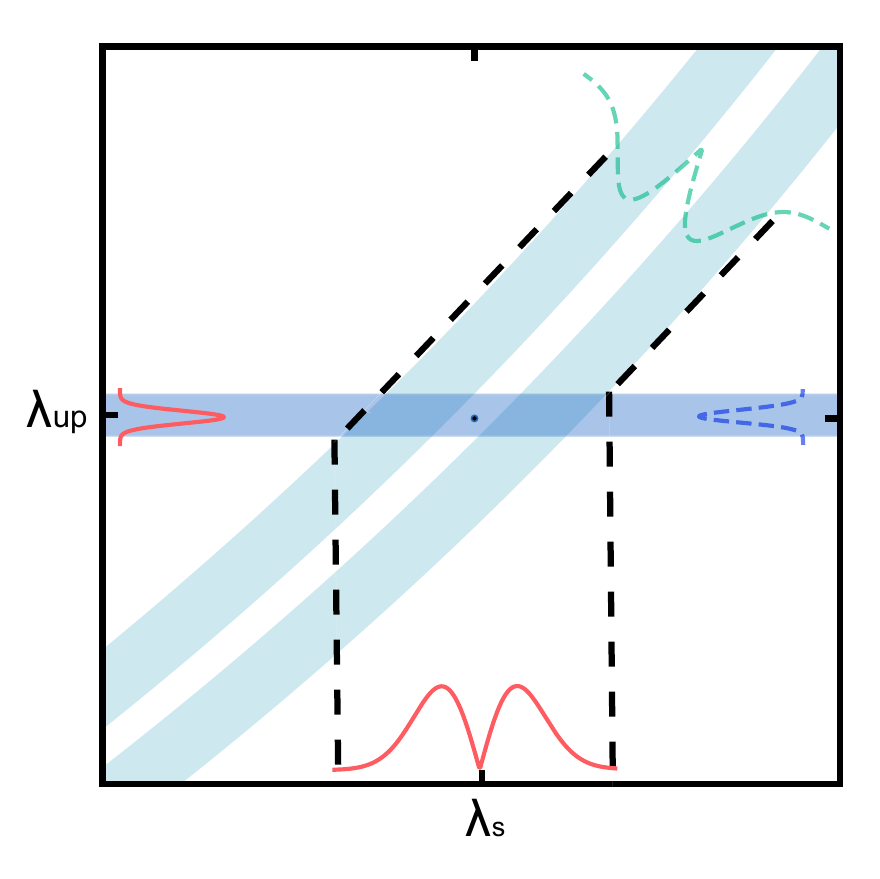}
  }
\caption{Schematic Transfer Function for mode-selective SPS. Here the incoming gate mode gets transferred onto the signal axis due to the intersection with the narrow, horizontal phase-matching function. There, the narrow and horizontal shape of the phase-matching function results in a narrow up-converted mode, suitable for heralding.}
  \label{fig:tf_schematics}
\end{figure}

To provide a clear schematic representation of the intended processes, we illustrate a target JSA for the SPA process in Fig.~\ref{fig:jsa_schematics} and a TF for the SPS process in Fig.~\ref{fig:tf_schematics}, each illustrating two different incoming classical modes. In both cases, the advantage of a horizontally oriented phase-matching function $\phi(\omega_i, \omega_k)$ becomes evident. When a strong, coherent beam — the pump in SPA or the gate in SPS — is present, its spectral shape $\alpha(\omega)$ is transferred onto the signal mode, enabling mode-selective heralded operations.
This highlights the potential of materials such as thin-film lithium niobate (TFLN), which enable dispersion engineering through geometrical tailoring. When combined with algorithmic optimization, these platforms provide a powerful route toward designing waveguides with desired spectral and quantum-optical properties.
In all configurations shown, the idler mode remains confined to a narrow, well-defined spectral region, facilitating a straightforward heralding scheme. However, due to the non-orthogonal geometric relation between the phase-matching function and the pump/gate envelope, the tilted overlap region induces a small deviation from perfect mode purity. This effect remains within an acceptable range, as discussed in Sec.~\ref{sec:results}.

\newpage
\subsection{Waveguide Simulator}
\label{sec:waveguide_simulator}
\subsubsection{Metallic Waveguide Approximation}
\label{sec:metallic_waveguide_approximation}

\noindent To put our optimization approach to the test, we begin by modeling the system using the metallic waveguide approximation described in this section. This model is fully analytically derivable and has been shown to yield good agreement between theoretical predictions and experimental results for diffused waveguides~\cite{RomanFainsin2024}.
We derive the refractive index and, subsequently, the phase-matching function used to construct the Joint Spectral Amplitudes (JSAs) and Transfer Functions (TFs). This approximation assumes perfectly rectangular waveguides with a discrete refractive index profile, confined strictly to the rectangular geometry of the waveguide (see Fig.~\ref{fig:waveguide_comparison}a).
The phase-matching function is approximated as

\begin{equation}
    \phi(\omega) = \text{sinc}\left(\frac{L}{2} \Delta k \right),
\end{equation}

where \( \Delta k = k_1 + k_2 - k_3 - \frac{2\pi}{\Lambda}\), and \( k(\omega) = \frac{n(\omega)\omega}{c} \) is the wave vector, with \( n(\omega) \) representing the refractive index as a function of frequency \cite{Boyd_Nonlinear_Optics_2008}. \(k_{1,2,3}\) therefore denote the wave vectors of each field involved in the nonlinear process. \(\Lambda\) is the poling period of the magnetic domains in the medium, which can be set during the fabrication process. To get an optimal efficiency, we aim for so-called perfect phase-matching, so \( \Delta k = 0 \). The poling period is typically chosen to satisfy this condition at the central frequencies of the interacting fields. The frequency-dependent refractive index \( n(\omega) \) is calculated using Sellmeier equations specific to the nonlinear material. For the metallic waveguide approximation, we use coefficients for periodically poled lithium niobate (pp:LN) and potassium titanyl phosphate (pp:KTP) as reported in~\cite{Edwards84,Jundt_97}.

\vspace{0.2cm}
\subsubsection{Thin-Film Model}
\label{sec:thinfilm_model}
\begin{figure}
    \centering
\includegraphics[width=0.9\textwidth]{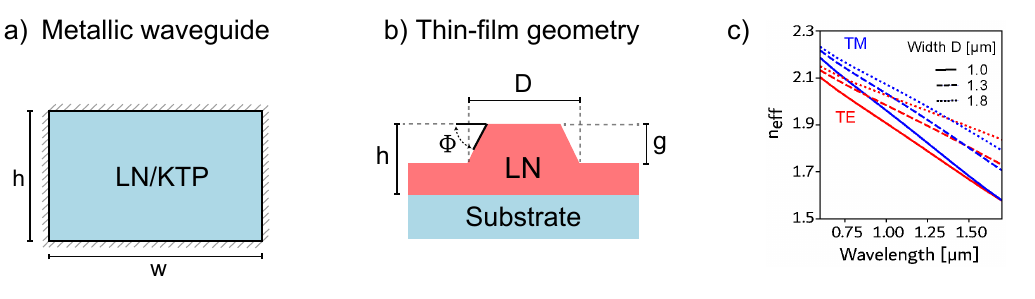}
\caption{Comparison of two waveguide geometries: (a) a metallic waveguide and (b) a thin-film waveguide. In (a), the cross-section is characterized by the height \( h \) and the width \( w \). In (b), the cross-section is characterized by the sidewall angle \( \phi \), the waveguide width \( D \), the etching depth \( g \), and the thin-film thickness \( h \). (c) Example values for $n_{eff}$ obtained for different TFLN waveguide geometries (besides D the parameters are equal to the ones reported in Table \ref{tab:SPA_thinfilm_addition}).}

    \label{fig:waveguide_comparison}
\end{figure}

\noindent Secondly, we investigate thin-film lithium niobate (TFLN) waveguides, where the effective refractive indices \(n_{\text{eff}}(\omega)\) of the guided modes strongly depend on the waveguide geometry. This high sensitivity to structural parameters provides a promising foundation for tailoring waveguides to our desired quantum processes by optimizing their geometrical structure \cite{jankowski_dispersion-engineered_2021, xin_spectrally_2022}.
Here, we consider trapezoidal waveguides as depicted in Fig. \ref{fig:waveguide_comparison}b, which are characterized by the thin film thickness h, the etching depth g, the waveguide width D and the side wall angle $\phi$. To obtain $n_{eff}$ we solve the wave equation utilizing the commercial software MODE: Waveguide Simulator \cite{Lumerical}, where we use  the refractive indices provided in \cite{Gayer2008} for lithium niobate and \cite{Palik1985} for the SiO2-Substrate. 

In the simulation results we identify the fundamental TE and TM modes, by searching for the modes with the highest $n_ {eff}$ with TE- and TM-fractions above 90$\%$. These $n_{eff}(\omega)$ profiles of the fundamental modes we then use to obtain the JSAs and TFs. Note, that we ignore geometries in which we identify mode crossings of the fundamental modes (TE-fractions below 90$\%$). 

Some example profiles for $n_{eff}$ which we obtain for different TFLN waveguide geometries are depicted in Fig.~\ref{fig:waveguide_comparison}c. These highlight the strong tunability of the dispersion properties via changes in the waveguides, which in the following will allow us to design SPA and SPS processes as discussed in Sec. 2.3.

\subsection{Purity and Mode-Selectivity}
\label{sec:purity}
For the sake of simplicity we consider the photon subtraction and addition processes on pure Gaussian multimode input states. Such states can always be written in some mode basis as \(\hat{\rho} = \ket{\phi_1}\bra{\phi_1} \otimes \ket{\phi_2}\bra{\phi_2} \otimes \cdots \otimes \ket{\phi_n}\bra{\phi_n}\). In temporal-spectral multimode Gaussian states, the modes in which the state is factorized is usually a Hermite Gauss basis \cite{RomanFainsin2024}. 

We are interested in subtracting or adding photons in a specific mode of the system, which, for simplicity can be just one of the Hermite Gauss modes, on which all the modes are disentangled. For this, it is essential to reduce the number of modes effectively involved in the non-linear interaction as close to 1 as possible. Since the JSA/TF of the process can be decomposed into a set of principal modes, the process itself can be characterized by the number of effective modes, similar to the Schmidt-number \cite{ekert1995}:
\[
K = \frac{\left( \sum_n \lambda_n \right)^2}{\sum_n \lambda_n^2}.
\]

This metric  is directly related to the output state's purity, where \(K = 1\) corresponds to a pure photon addition or subtraction from a pure input. It should be noted that when we refer to the Schmidt number and to the purity, we refer to the purity of the process itself, and therefore relate to an outcome multimode-state from an pure incoming state. 

In case of multimode processes, we can express the output state of the single photon subtraction process as
    
\begin{equation}
    \hat{\rho}_s^{\text{out}} = \sum_{l} \tilde{\lambda}_l \, \hat{A}_l \, \hat{\rho}_s^{\text{in}} \, \hat{A}_l^\dagger
\end{equation}

where \(\tilde{\lambda}_l = \lambda_l / P^-\) are the normalized squared eigenvalues of the decomposed TF, weighting  the contribution of the specific mode with ladder-operators $\{\hat{A}_l, \hat{A}^\dagger_l\}$. The normalization constant, $P^- = \sum_{l,m} \lambda_l O_{lm}\bar{n_m}$, defines the probability of success of the heralding detection. Here $\bar n_m$ defines the average photon number in the $m^{th}$ input mode, and $O_{lm}$ are the coefficients in the decomposition of the $l^{th}$ eigenmode of the process ($\hat A_l$), in terms of the input modes ($\hat a_m$), $\hat A_l=\sum_m O_{lm} \hat a_m$.\\
The purity of the output state, which, given that the input state is pure, is directly linked to the mode selectivity of the photon subtraction process is thus given by [see Appendix]:

\[
\mu^-=\text{Tr}(\hat{\rho}_{out}^2) = \sum_{l,k}\tilde\lambda_l \tilde\lambda_k \left(\sum_m O_{lm} O_{km} \bar n_m\right)^2 .\] 
In the case in which $K=1$ and only a single mode is involved in the non-linear process the expression above evaluates exactly to one. It is also possible to notice that if only one of the input modes has non zero average photon number then the purity is also one, as all other modes do not contribute, with the caveat that the output mode gets projected into this single mode. \\
Similarly, the output for a SPA-process can be written as:
\begin{equation}
    \hat{\rho}_s^{\text{out}} = \sum_{l} \tilde{\lambda}_l \hat{A}_l^\dagger \hat{\rho}_s^{\text{in}} \hat{A}_l
\end{equation}
Using the same procedure as above, one can arrive at:

\begin{equation}
\mu^+ = \sum_{l,k}\tilde \lambda_l \tilde\lambda_k \left | \sum_m O_{lm}O_{km}(1+\bar n_m)\right|^2,
\end{equation}

where \(\bar{n}_k\) is the mean photon number in the input mode k. Differently to the case of photon subtraction, here all input modes, including those in vacuum, contribute to the final state. This implies that in the case of photon addition, only by having exactly $K=1$ the process can produce a pure outcome state.
\\
These results provide a framework to understand the connection between the Schmidt eigenvalues and the mode selectivity of the process.
The purity \( \mu = 1 \) is attainable for both processes only when the Schmidt decomposition reduces to a single non-zero eigenvalue \( \lambda \), corresponding to a single involved mode in the signal and idler fields. This condition requires symmetry along the signal and idler axes in the Joint Spectral Amplitude (JSA), as outlined in \cite{Grice_walmsley_2001}. However, in our experimental setup, such symmetry is inherently constrained by the interplay between the waveguide's phase-matching function and the pump spectral profile [see Fig.~\ref{fig:tf_metallic_group}, \ref{fig:modes_SPS_metallic_group}, \ref{fig:tf_thinfilm_group}]. Despite this limitation, approximating this symmetry remains desirable for our objectives. 
While achieving a Schmidt number of \( K = 1 \) is theoretically unattainable, a value close to unity remains our focus. Therefore, we direct our search within the parameter space towards configurations approaching \( K \approx 1 \). Our primary goal is to design a system that supports a narrow idler-frequency bandwidth for the heralding scheme, while fully exploiting the pump bandwidth in the signal state. To achieve this, we make use of the narrow phase-matching width along the idler axis within our target frequency range, as discussed in Sec.~2.3.\\
An additional method that can be straightforwardly applied to any outcome of our optimization should be further mentioned — namely, \textit{spectral filtering along the idler axis}. This can be experimentally realized using a narrowband optical filter or, when fine spectral selectivity is required, a dispersive element, e.g., a diffraction grating, combined with a slit, to isolate a specific spectral window. The filtering is implemented between the photon generation and the heralding process. While it reduces the overall heralding rate by discarding undesired frequency components, it does not degrade the quantum properties of the heralded state or the purity of the process. With this technique, the process can asymptotically approach unit purity while decreasing the event rate of the process~\cite{Raymer2005}.

\subsection{Optimization and Numerical Search}
Since we have demonstrated in Sec.~2.5 that achieving a Schmidt number of \( K \approx 1 \), together with the appropriate shape of the phase-matching function in the LN and KTP waveguides, is sufficient to configure the idler as a narrow-bandwidth heralding mode while transferring the pump bandwidth to the broader signal field, we proceed with a numerical optimization approach, described in Methods. To apply our algorithm to a fully analytical testbed, we model the nonlinear waveguide using the metallic-waveguide approximation, incorporate periodic poling for quasi-phase matching, and determine the refractive index via the Sellmeier equations introduced in Sec.~2.4.a. The phase-matching function is computed based on the waveguide geometry and the type of nonlinear process (Type~0, I, or II), all treated as variable parameters. In addition, we explore the influence of the Hermite–Gauss pump mode order and bandwidth on the resulting JSAs and transfer functions. Using a genetic algorithm, we perform a parameter-space search to minimize the Schmidt number toward \( K \approx 1 \), thereby identifying waveguide configurations suitable for mode-selective single-photon addition and subtraction.

\subsection{Testbed Results - Metallic Waveguide approximation}
\label{sec:results_testbed_met}
As mentioned in Sec.~2.4.a, we begin by implementing our scheme within the metallic waveguide approximation. Within this framework, we present results that already demonstrate systems capable of performing mode-selective non-Gaussian operations. In cases where differences in mode selectivity were minimal, we opted for the marginally larger pump width to ensure easier experimental implementation in terms of pulse shaping.
\subsubsection{Mode-Selective Single-Photon Addition}
\label{sec:SPA_metallic}
\begin{figure}[H]
    \centering
    \subfloat[Pump mode: HG\(_0\)]{%
        \includegraphics[width=0.33\textwidth]{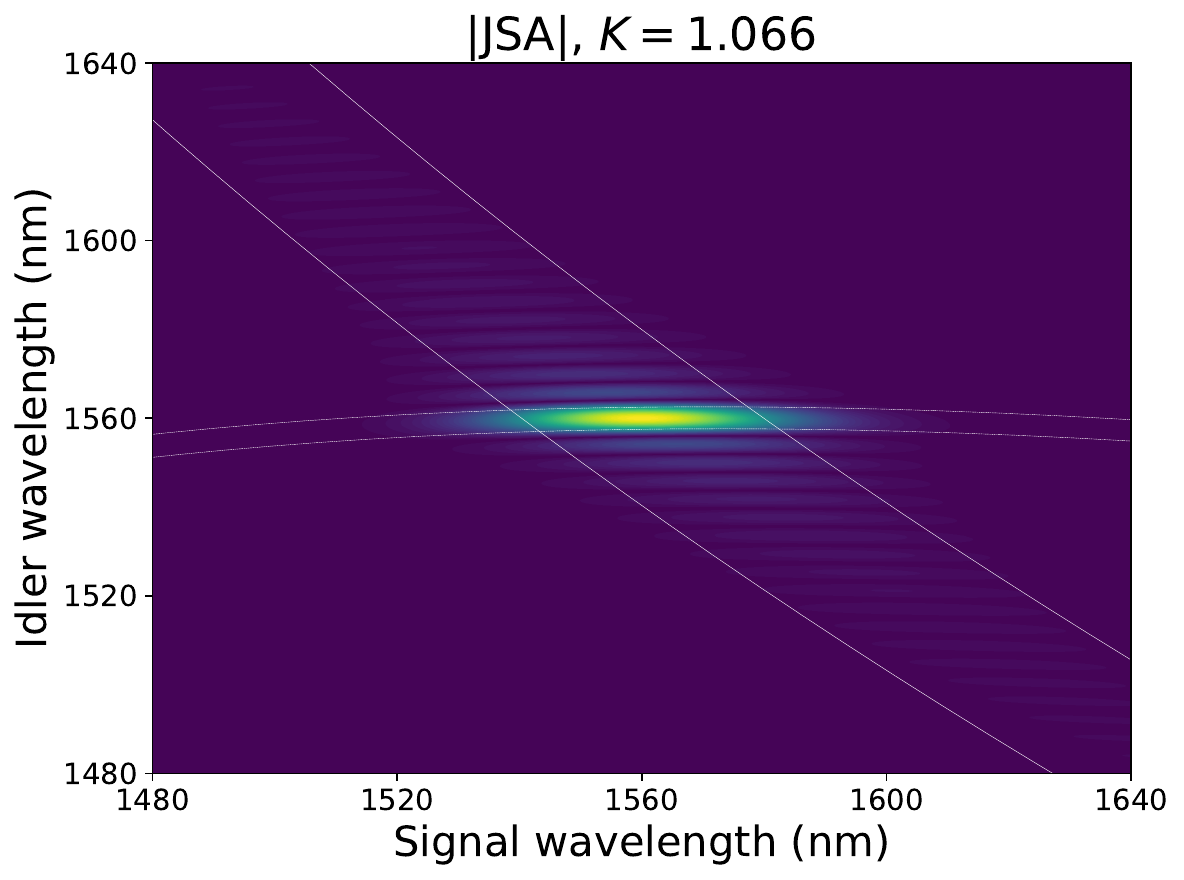}
        \label{fig:jsa_hg0}
    }
    \subfloat[Pump mode: HG\(_1\)]{%
        \includegraphics[width=0.33\textwidth]{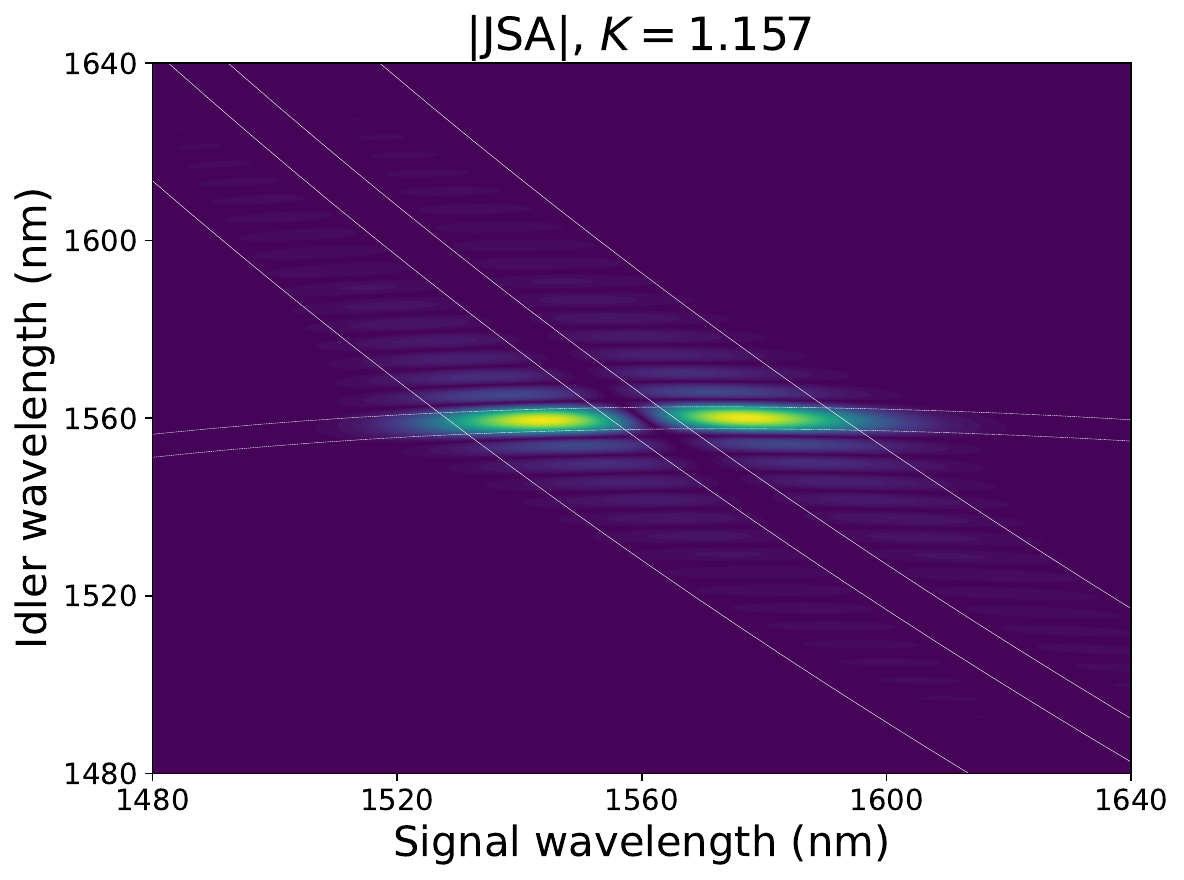}
        \label{fig:jsa_hg1}
    }
    \subfloat[Pump mode: HG\(_2\)]{%
        \includegraphics[width=0.33\textwidth]{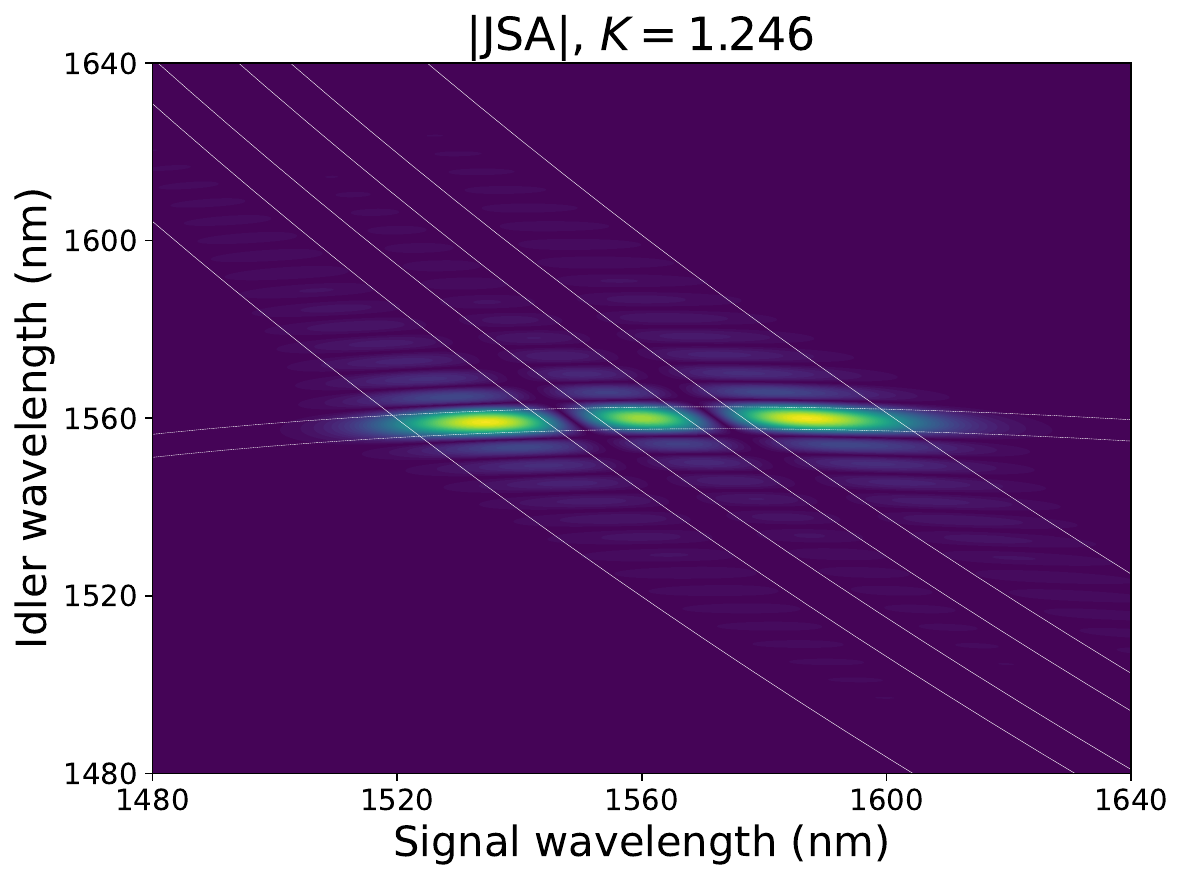}
        \label{fig:jsa_hg2}
    }

    \caption{Joint Spectral Amplitudes (JSAs) corresponding to different pump modes, calculated using the waveguide and pump parameters listed in Table~\ref{tab:waveguide_2}. The involved frequency ranges of the signal and idler modes are directly indicated. Via a diagonalization similar to the Schmidt decomposition, the involved eigenmodes characterizing the purity of the process can be obtained and are shown in Fig.~\ref{fig:SPA_metallic_modes_group}.}
    \label{fig:jsa_group}
\end{figure}

\begin{figure}[H]
    \centering
    \subfloat[Obtained signal mode with pump: HG\(_0\)]{%
        \includegraphics[width=0.48\textwidth]{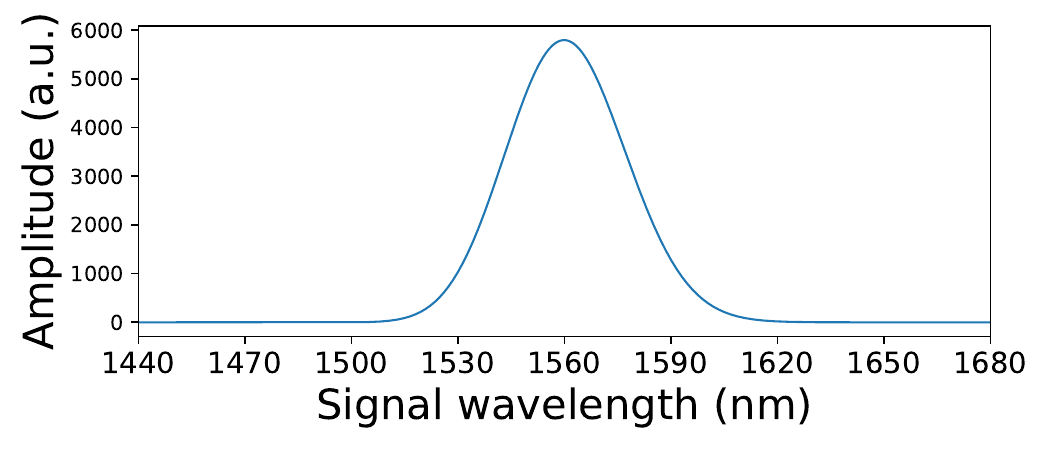}
        \label{fig:spa_met_modes_hg0}
    } \hfill
    \subfloat[Obtained signal mode with pump: HG\(_1\)]{%
        \includegraphics[width=0.48\textwidth]{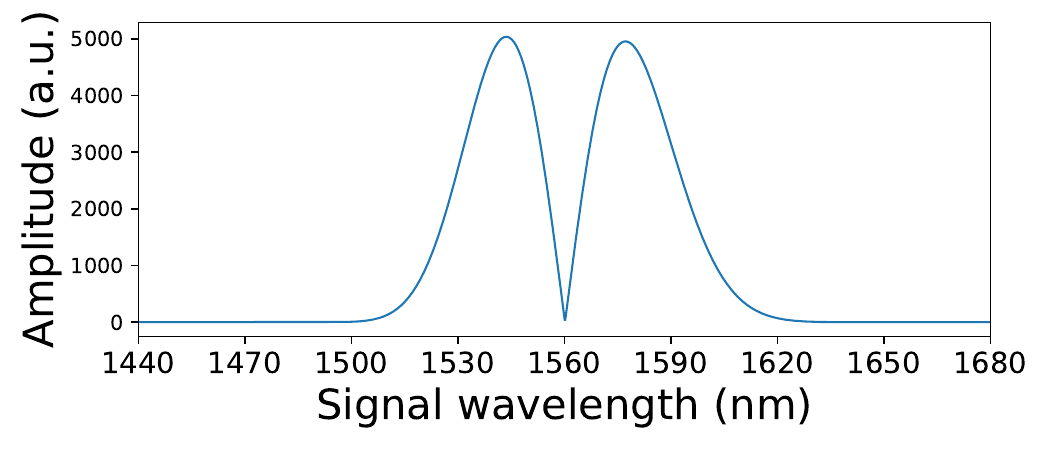}
        \label{fig:spa_met_modes_hg1}
    } \vspace{0.0cm} 
    \subfloat[Obtained signal mode with pump: HG\(_2\)]{%
        \includegraphics[width=0.48\textwidth]{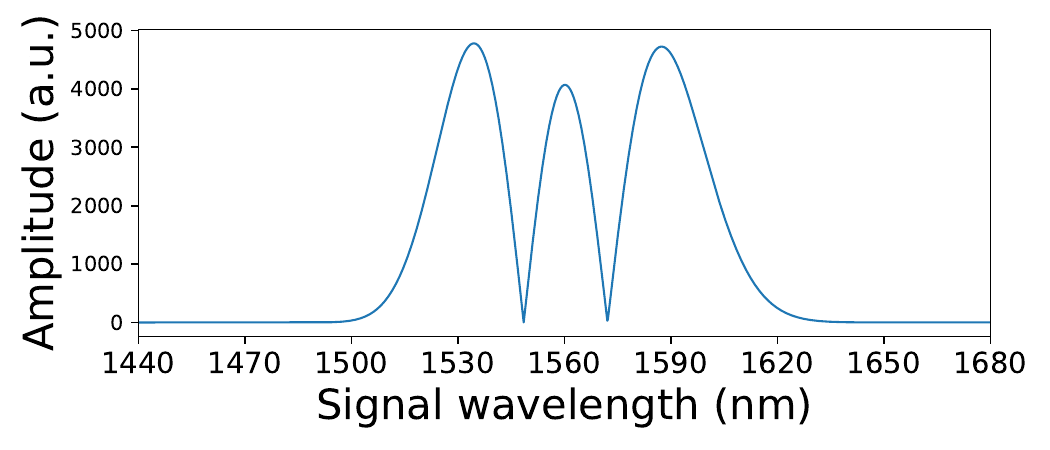}
        \label{fig:spa_met_modes_hg2}
    } \hfill
    \subfloat[Obtained idler mode. Obtained with HG\(_0\)-mode as pump. Representative for all pump-modes]{%
        \includegraphics[width=0.48\textwidth]{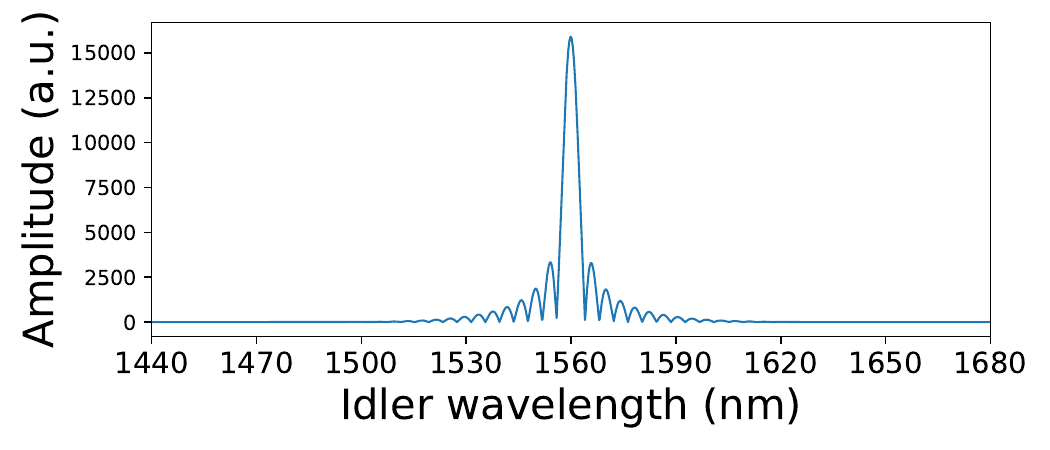}
        \label{fig:spa_met_modes_idler}
    }
    \caption{Signal and idler modes corresponding to the JSAs shown in Fig.~\ref{fig:jsa_group}, obtained using the parameters listed in Table~\ref{tab:waveguide_2}.}
    \label{fig:SPA_metallic_modes_group}
\end{figure}

\noindent We start by showing the results for the SPA scheme. We identified an optimized waveguide-pump configuration, with the corresponding parameters listed in Table~\ref{tab:waveguide_2}. These parameters yield the desired JSAs and spectral modes, shown in Figs.~\ref{fig:jsa_group} and \ref{fig:SPA_metallic_modes_group}.
Fig.~\ref{fig:jsa_group} displays the JSAs for Hermite-Gaussian (HG) pump modes of orders 0, 1, and 2 - all for the same waveguide. Correspondingly, Fig.~\ref{fig:SPA_metallic_modes_group} illustrates the signal and idler modes obtained from the JSA decomposition. The signal modes span over approximately 100~nm, while the idler mode serves as the desired narrow-band mode, enabling an efficient heralding detection scheme.
Since the idler mode remains nearly unchanged across all HG orders, we present only the zero-order mode as a representative example for all idler modes.

\begin{table}[H]
    \centering
    \small
    \subfloat[Waveguide and pump parameters used to generate the JSAs for the SPA process and the associated modes shown in Figs.~\ref{fig:jsa_group} and \ref{fig:SPA_metallic_modes_group}.]{%
        \begin{tabular}{|l|l|}
        \hline
        \textbf{Parameter}             & \textbf{Value}             \\ \hline
        Length                         & 7 mm                       \\ \hline
        Width                          & 2.8 $\mu$m                 \\ \hline
        Height                         & 2.3 $\mu$m                 \\ \hline
        Material                       & KTP                        \\ \hline
        Pump width                     & 7 nm                     \\ \hline
        Pump mode                      & HG 0/1/2                   \\ \hline
        Type                           & II                         \\ \hline
        Polarization (signal)          & H                          \\ \hline
        Polarization (idler)           & V                          \\ \hline
        Polarization (pump)            & H                          \\ \hline
        \end{tabular}
        \label{tab:waveguide_2}
    } \hspace{8em}
    \subfloat[Waveguide and pump parameters used to generate the TFs for the SPS process and the associated modes shown in Figs.~\ref{fig:tf_metallic_group} and \ref{fig:modes_SPS_metallic_group}.]{%
        \begin{tabular}{|l|l|}
        \hline
        \textbf{Parameter}             & \textbf{Value}             \\ \hline
        Length                         & 2.0 mm                     \\ \hline
        Width                          & 2.5 $\mu$m                 \\ \hline
        Height                         & 1.8 $\mu$m                 \\ \hline
        Material                       & KTP                        \\ \hline
        Pump width                     & 10 nm                      \\ \hline
        Pump mode                      & HG 0/1/2                   \\ \hline
        Type                           & 0                          \\ \hline
        Polarization (signal)          & V                          \\ \hline
        Polarization (idler)           & V                          \\ \hline
        Polarization (pump)            & V                          \\ \hline
        \end{tabular}
        \label{tab:parameters_SPS_150runs_triple}
    }
    \caption{Waveguide and pump parameters obtained under the metallic waveguide approximation for a mode-selective SPA and SPS process.}
    \label{tab:waveguide_comparison}
\end{table}

\subsubsection{Mode-Selective Single-Photon Subtraction}
\begin{figure}[H]
    \centering
    \subfloat[Gate mode: HG\(_0\)]{%
        \includegraphics[width=0.32\textwidth]{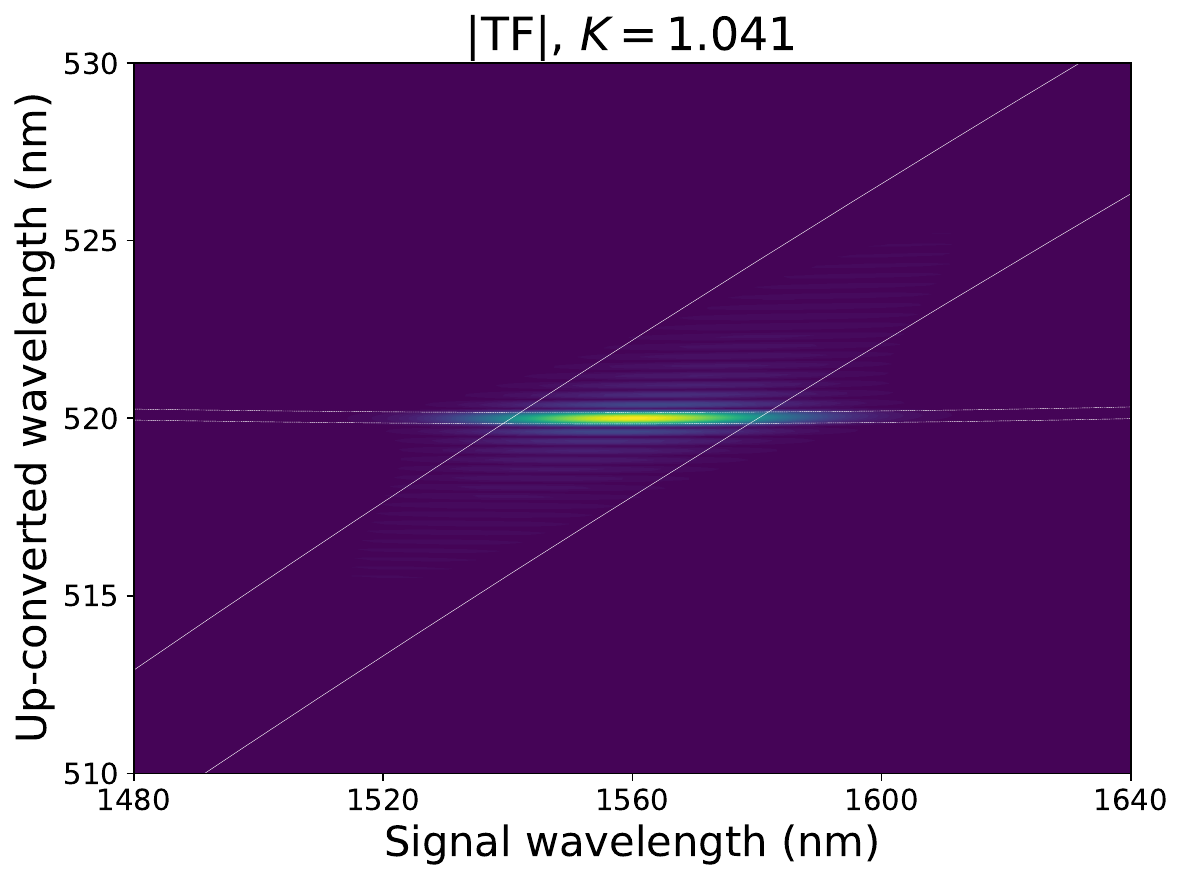}
        \label{fig:met_tf_hg0}
    }
    \subfloat[Gate mode: HG\(_1\)]{%
        \includegraphics[width=0.32\textwidth]{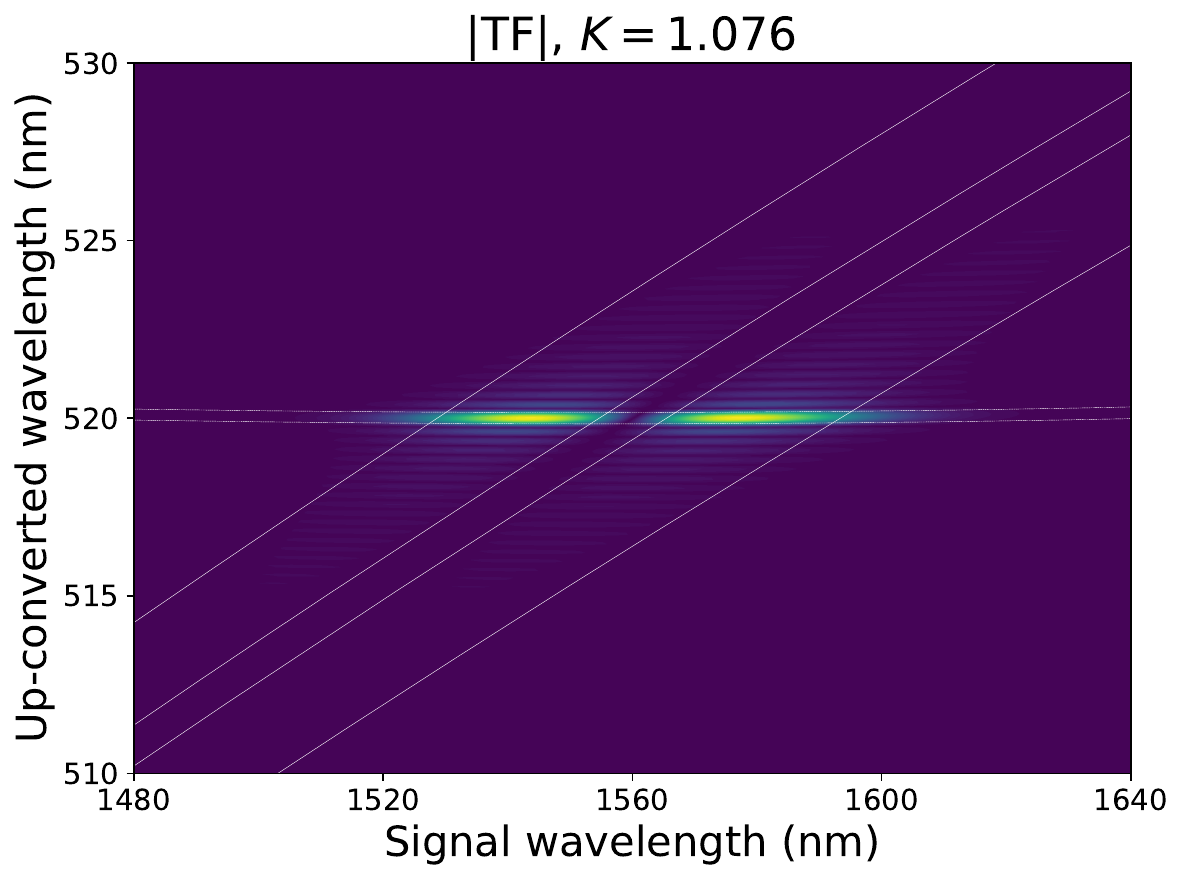}
        \label{fig:met_tf_hg1}
    }
    \subfloat[Gate mode: HG\(_2\)]{%
        \includegraphics[width=0.32\textwidth]{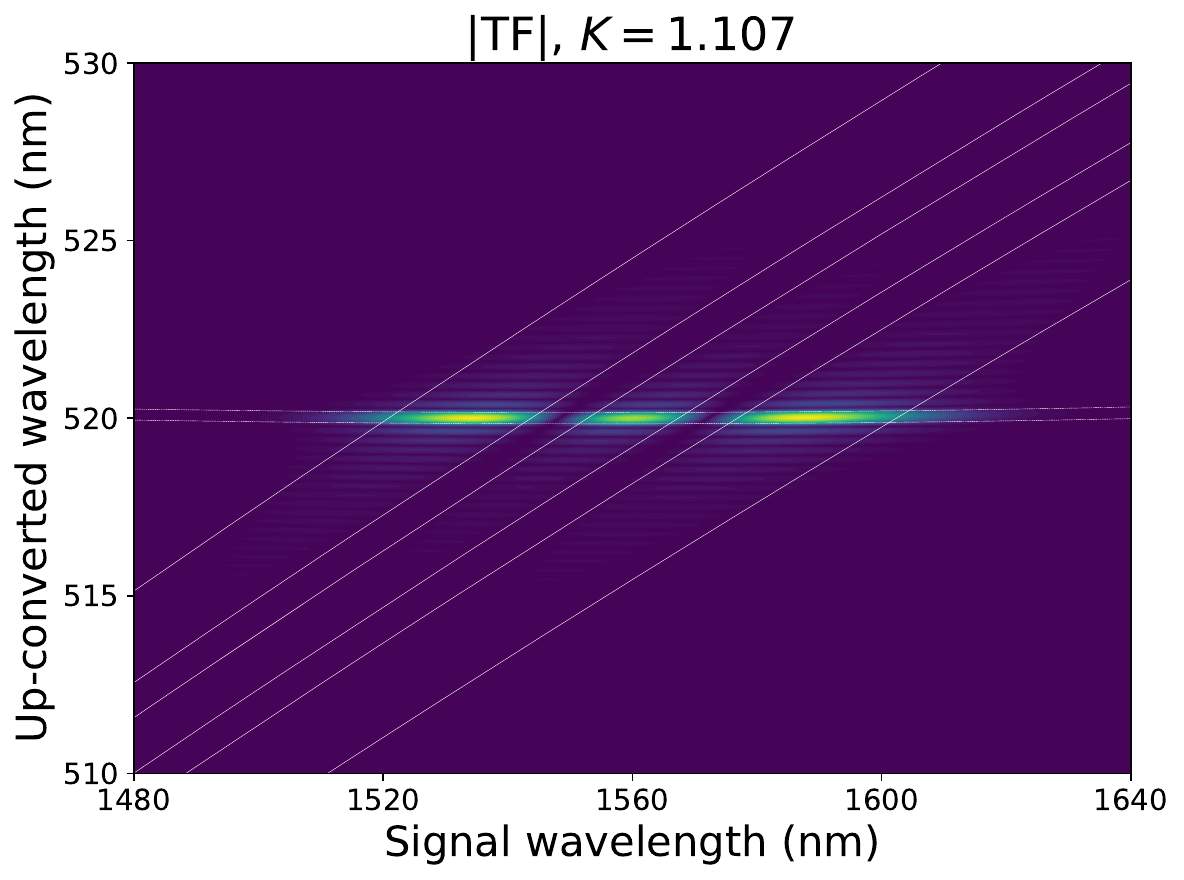}
        \label{fig:met_tf_hg2}
    }

    \caption{Transfer functions (TFs) corresponding to the pump modes HG\(_0\), HG\(_1\), and HG\(_2\), calculated using the parameters listed in Table~\ref{tab:parameters_SPS_150runs_triple}. The involved frequency ranges of the signal and up-converted modes are directly indicated. }
    \label{fig:tf_metallic_group}
\end{figure}

\begin{figure}[H]
    \centering
    \subfloat[Obtained signal mode with pump: HG\(_0\)]{%
        \includegraphics[width=0.48\textwidth]{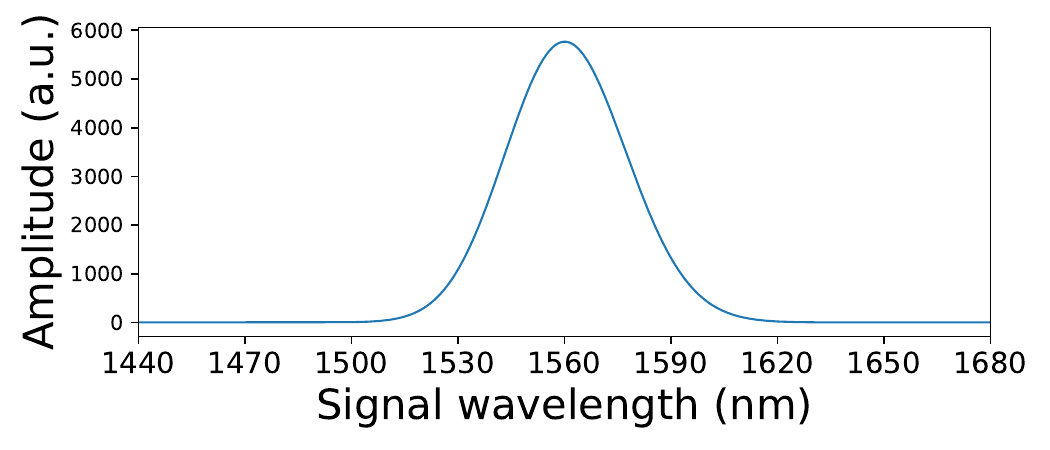}
        \label{fig:sps_met_modes_hg0}
    } \hfill
    \subfloat[Obtained signal mode with pump: HG\(_1\)]{%
        \includegraphics[width=0.48\textwidth]{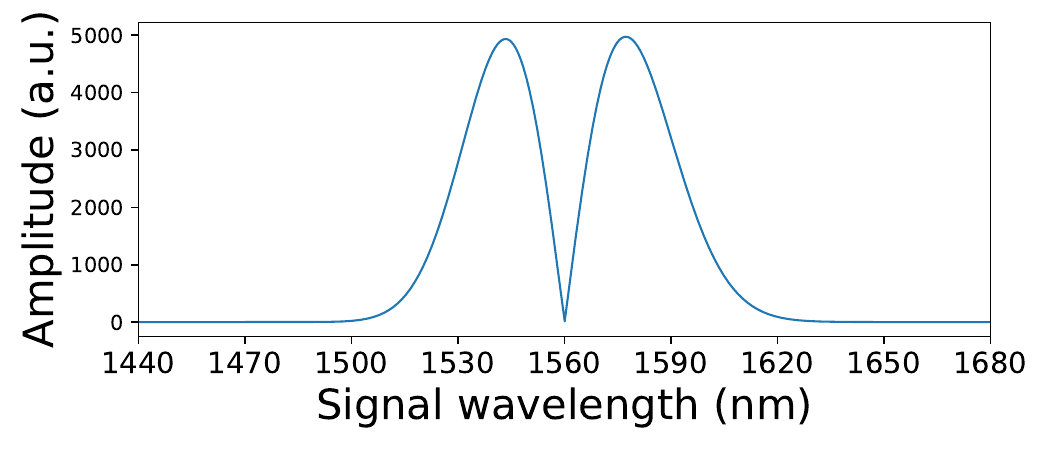}
        \label{fig:sps_met_modes_hg1}
    } \vspace{0.2cm} 
    \subfloat[Obtained signal mode with pump: HG\(_2\)]{%
        \includegraphics[width=0.48\textwidth]{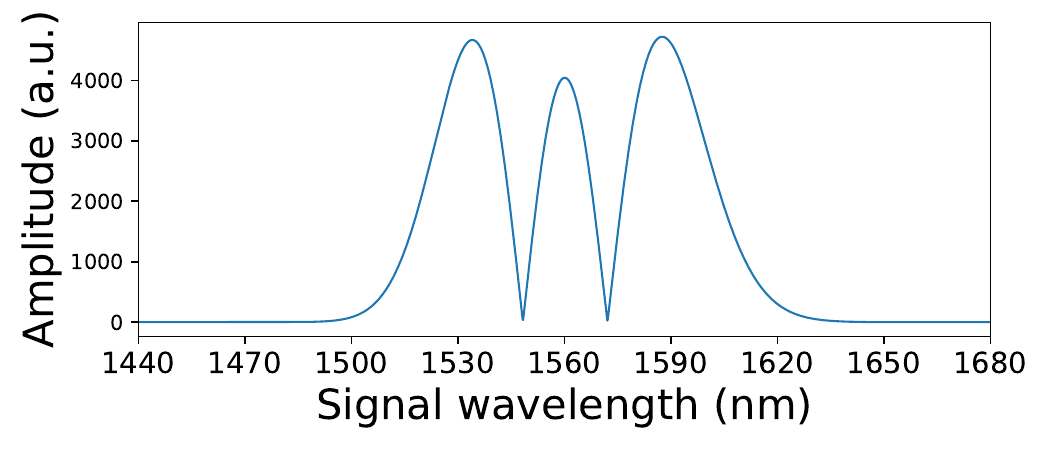}
        \label{fig:sps_met_modes_hg2}
    } \hfill
    \subfloat[Obtained gate mode for HG\(_0\)-pump. Representative for all gate-modes.]{%
        \includegraphics[width=0.48\textwidth]{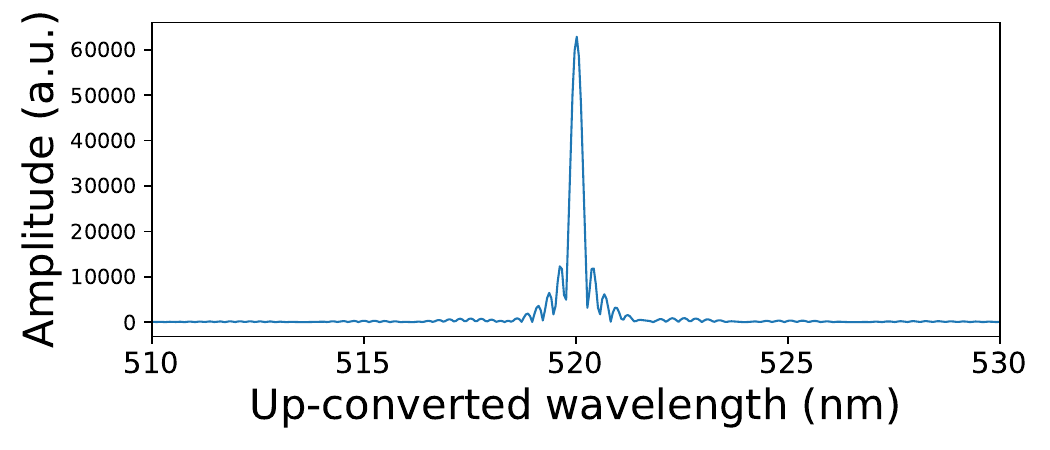}
        \label{fig:sps_met_modes_idler}
    }
    \caption{Signal and gate modes corresponding to the TFs shown in Fig.~\ref{fig:tf_metallic_group}, obtained using the parameters listed in Table~\ref{tab:parameters_SPS_150runs_triple}.}
    \label{fig:modes_SPS_metallic_group}
\end{figure}

\noindent We further could identify a sufficient waveguide-pump-configuration for the SPS-process. The parameters are listed in Table~\ref{tab:parameters_SPS_150runs_triple} and yield the desired transfer functions (TFs) and spectral modes, shown in Figs.~\ref{fig:tf_metallic_group} and \ref{fig:modes_SPS_metallic_group}.  
Fig.~\ref{fig:tf_metallic_group} displays the TFs for Hermite-Gaussian gate modes of orders 0, 1, and 2. Correspondingly, Fig.~\ref{fig:modes_SPS_metallic_group} illustrates the signal and up-converted modes obtained from the TF decomposition. The signal modes span approximately 110~nm, while the up-converted mode serves as the desired narrow-band mode, enabling an efficient heralding detection scheme.  
Since the gate mode remains nearly unchanged across all HG orders, we present only the zero-order mode as a representative example for all up-converted modes.

\section{Results - Thin-Film Lithium Niobate waveguides}
\label{sec:results}
Since we could apply our optimization scheme to the analytical model, we now present our implementation to the TFLN-model (see Sec.~2.4.b), showing how finely tuned the dispersion engineering can be done to achieve the desired non-gaussian processes. 

\subsection{Single-Photon Addition}
\begin{figure}[H]
    \centering
    \subfloat[Pump mode: HG\(_0\)]{%
        \includegraphics[width=0.31\textwidth]{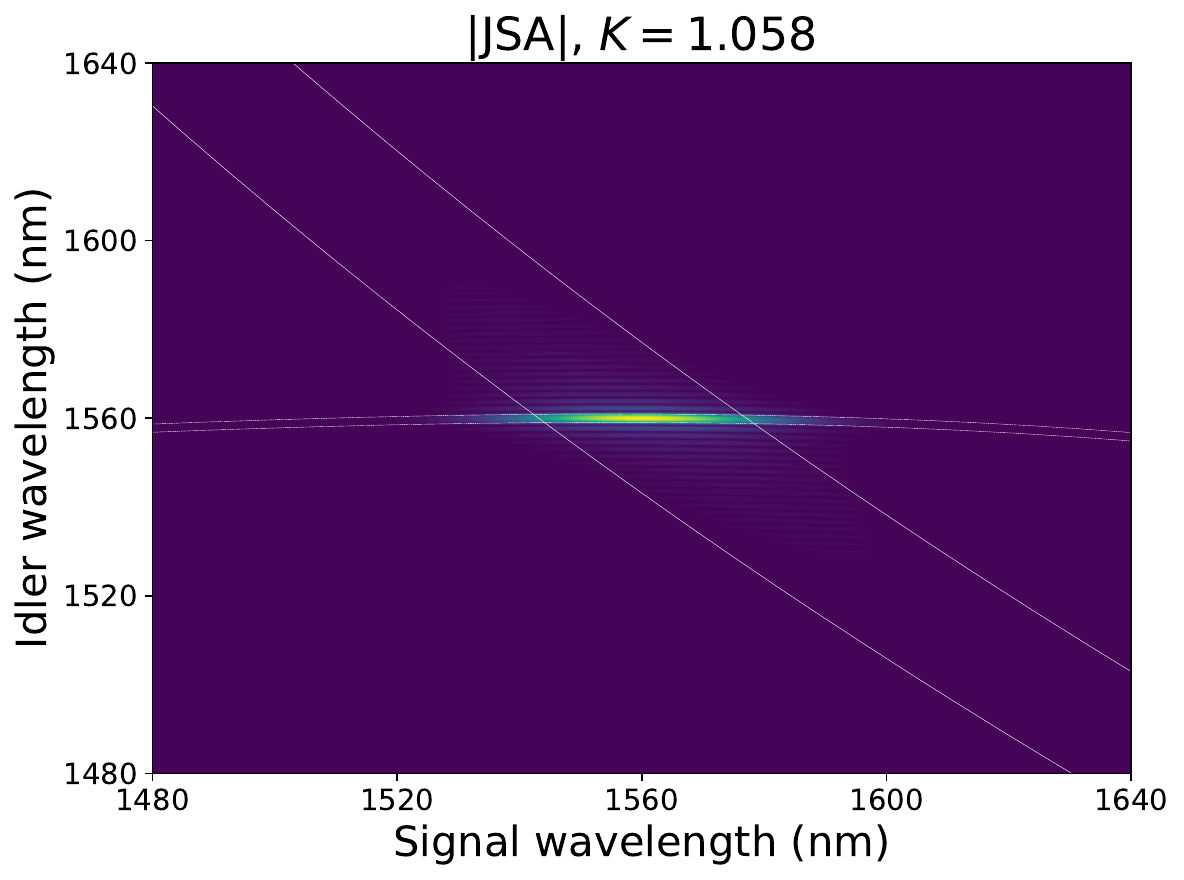}
        \label{fig:jsa_hg0_thinfilm}
    } \hfill
    \subfloat[Pump mode: HG\(_1\)]{%
        \includegraphics[width=0.31\textwidth]{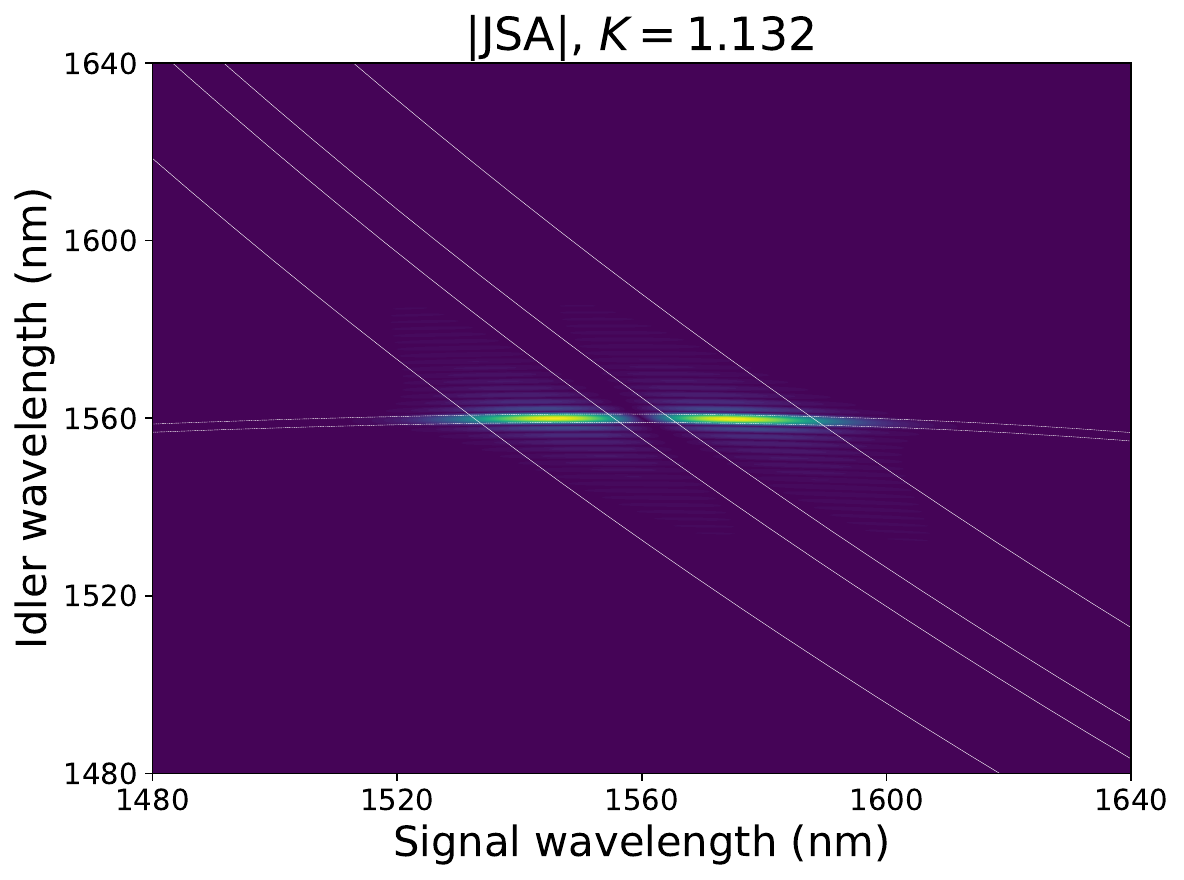}
        \label{fig:jsa_hg1_thinfilm}
    } \hfill
    \subfloat[Pump mode: HG\(_2\)]{%
        \includegraphics[width=0.31\textwidth]{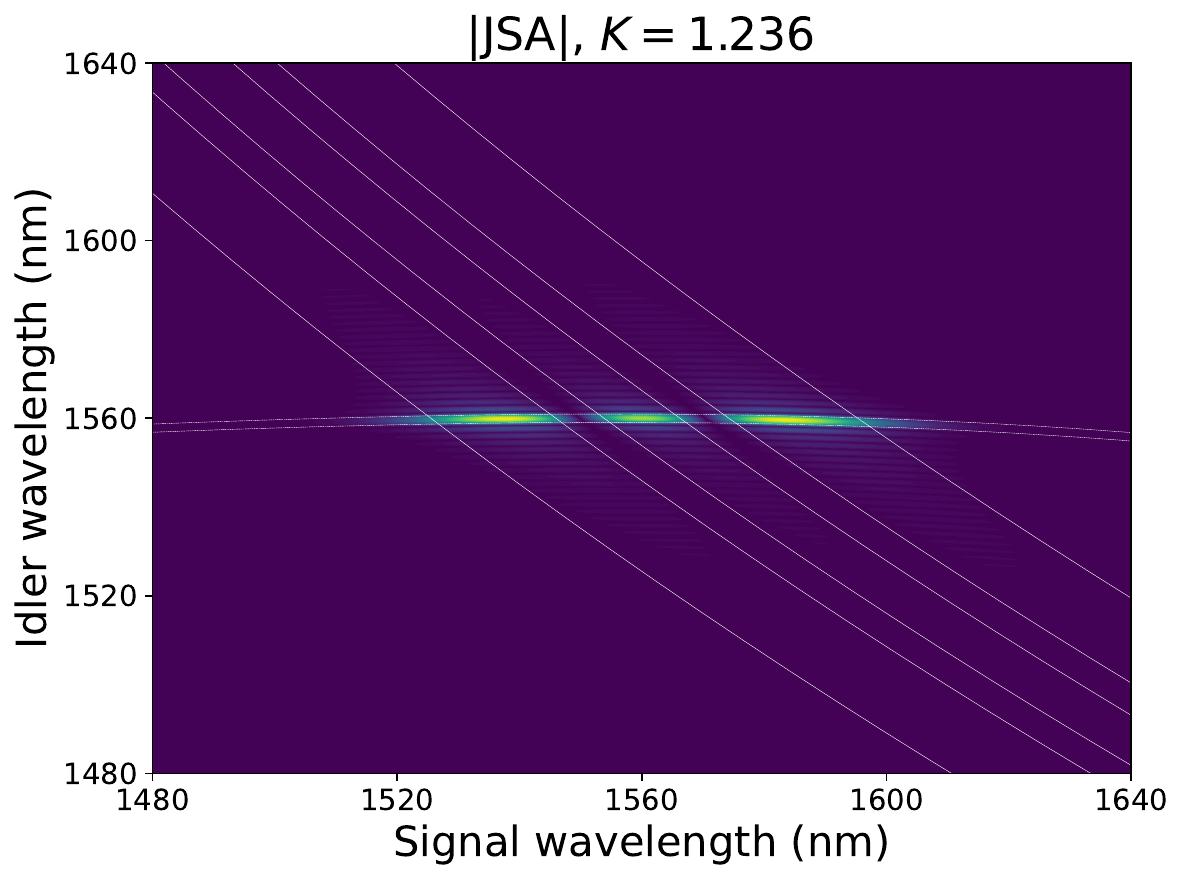}
        \label{fig:jsa_hg2_thinfilm}
    }

    \caption{Joint-Spectral Amplitudes (JSA) for HG\(_0\), HG\(_1\), and HG\(_2\), corresponding to the parameters in Table~\ref{tab:SPA_thinfilm_addition}. As in the metallic approximation, the involved quantum modes can be obtained by a Schmidt-decomposition and are presented in Fig. \ref{fig:modes_SPA_thinfilm_group}.}
    \label{fig:jsa_thinfilm_group}
\end{figure}

\begin{figure}[H]
    \centering
    \subfloat[Obtained signal mode with pump: HG\(_0\)]{%
        \includegraphics[width=0.48\textwidth]{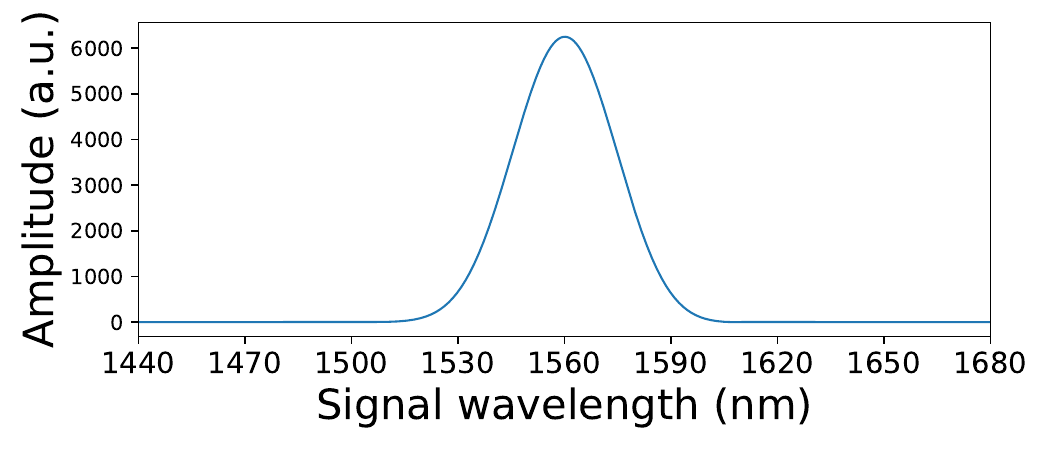}
        \label{fig:modes_spa_hg0_thinfilm}
    } \hfill
    \subfloat[Obtained signal mode with pump: HG\(_1\)]{%
        \includegraphics[width=0.48\textwidth]{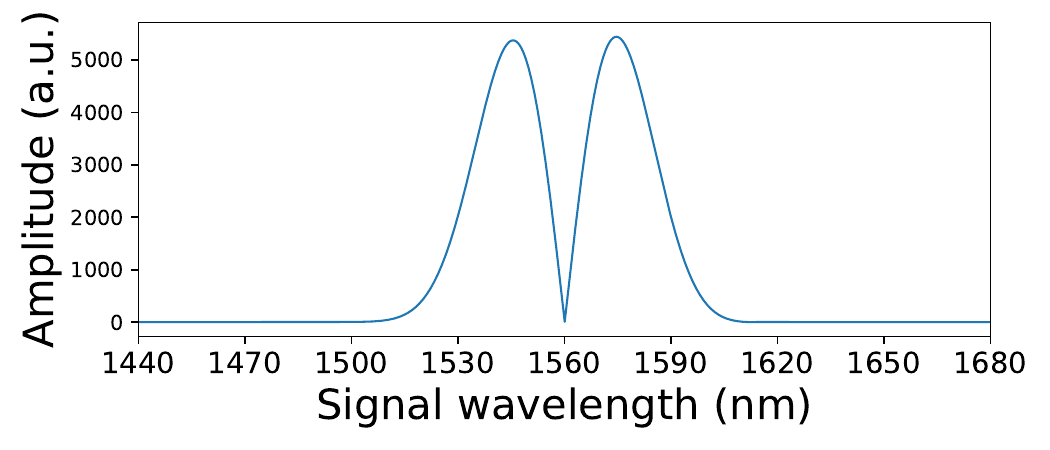}
        \label{fig:modes_spa_hg1_thinfilm}
    } \vspace{0.5cm} 
    \subfloat[Obtained signal mode with pump: HG\(_2\)]{%
        \includegraphics[width=0.48\textwidth]{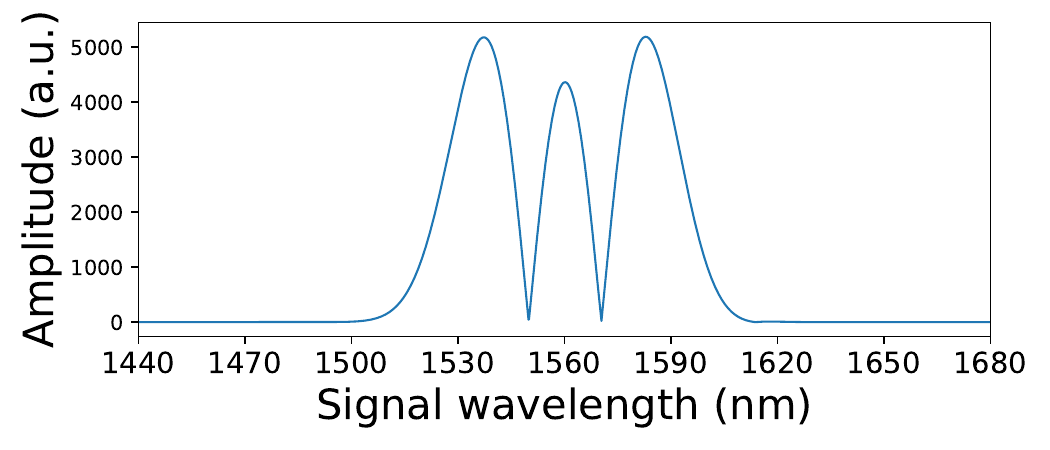}
        \label{fig:modes_hg2_thinfilm}
    } \hfill
    \subfloat[Obtained idler mode for HG\(_0\)-pump. Representative for all gate modes.]{%
        \includegraphics[width=0.48\textwidth]{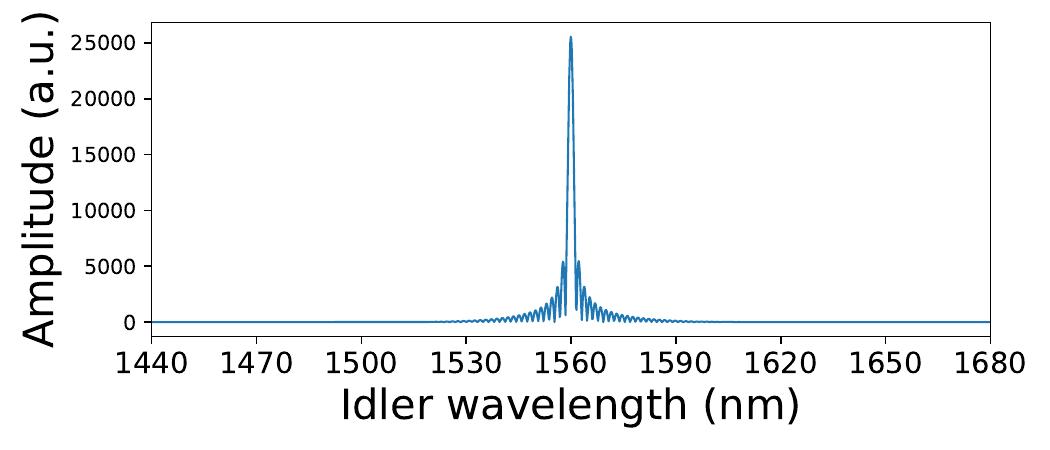}
        \label{fig:modes_spa_idler_thinfilm}
    }
    \caption{Signal and idler modes corresponding to the JSAs shown in Fig.~\ref{fig:jsa_thinfilm_group}, obtained using the parameters listed in Table~\ref{tab:SPA_thinfilm_addition}.}
    \label{fig:modes_SPA_thinfilm_group}
\end{figure}

Starting from the thin-film model, we identified an optimized waveguide-pump configuration for mode-selective single-photon addition. The respective waveguide- and pump-parameters are listed in Table~\ref{tab:SPA_thinfilm_addition}. The obtained Joint Spectral Amplitudes (JSAs) for HG pump modes of orders 0, 1, and 2 are shown in Fig.~\ref{fig:jsa_thinfilm_group}, while the corresponding signal and idler modes, extracted from the JSA decomposition, are presented in Fig.~\ref{fig:modes_SPA_thinfilm_group}. Since the idler mode remains nearly unchanged across all HG orders, we present only the zero-order signal mode as a representative example.

\begin{table}
    \centering
    \small
    \subfloat[Waveguide and pump parameters used to generate the JSAs for the SPA process and the associated modes shown in Figs.~\ref{fig:jsa_thinfilm_group} and \ref{fig:modes_SPA_thinfilm_group}.]{%
        \begin{tabular}{|l|l|}
        \hline
        \textbf{Parameter}             & \textbf{Value}             \\ \hline
        Length                         & 7 mm                       \\ \hline
        Width / D                      & 608 nm                    \\ \hline
        Height / h                     & 600 nm                     \\ \hline
        Etching angle / \(\phi\)       & 78.2°                      \\ \hline
        Etching depth / g              & 441 nm                     \\ \hline
        Material                       & LN                         \\ \hline
        Pump width                     & 6 mm                       \\ \hline
        Pump mode                      & HG 0/1/2                   \\ \hline
        Type                           & II                         \\ \hline
        Polarization (signal)          & V                          \\ \hline
        Polarization (idler)           & H                          \\ \hline
        Polarization (pump)            & V                          \\ \hline
        \end{tabular}
        \label{tab:SPA_thinfilm_addition}
    } \hspace{8em}
    \subfloat[Waveguide and pump parameters used to generate the TFs for the SPS process and the associated modes shown in Figs.~\ref{fig:tf_thinfilm_group} and \ref{fig:modes_SPS_thinfilm_group}.]{%
        \begin{tabular}{|l|l|}
        \hline
        \textbf{Parameter}             & \textbf{Value}             \\ \hline
        Length                         & 7 mm                       \\ \hline
        Width / D                      & 883 nm                     \\ \hline
        Height / h                     & 788 nm                     \\ \hline
        Etching angle / \(\phi\)       & 75.2°                      \\ \hline
        Etching depth/ g               & 774 nm                     \\ \hline
        Material                       & LN                         \\ \hline
        Pump width                     & 7 mm                       \\ \hline
        Pump mode                      & HG 0/1/2                   \\ \hline
        Type                           & II                         \\ \hline
        Polarization (signal)          & H                          \\ \hline
        Polarization (idler)           & V                          \\ \hline
        Polarization (pump)            & V                          \\ \hline
        \end{tabular}
        \label{tab:SPS_thinfilm_subtraction}
    }
    \caption{Comparison of waveguide and pump parameters for thin-film configurations enabling single-photon subtraction and addition. The parameters width (\(D\)), height (\(h\)), etching angle (\(\phi\)), and etching depth (\(g\)) are illustrated in Fig.~\ref{fig:waveguide_comparison}b.}
    \label{tab:SPS_thinfilm_comparison}
\end{table}

\subsection{Single-Photon Subtraction}

\begin{figure}[H]
    \centering
    \subfloat[Pump mode: HG\(_0\)]{%
        \includegraphics[width=0.31\textwidth]{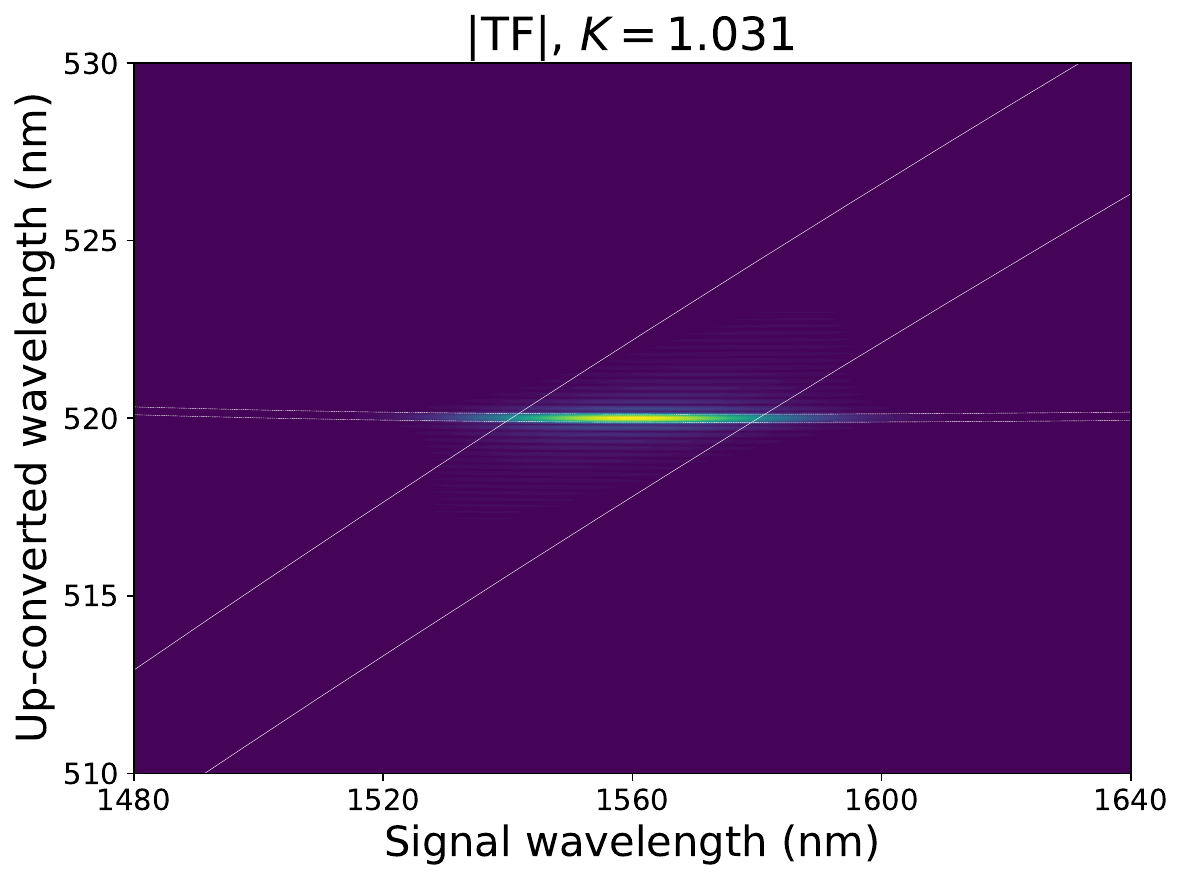}
        \label{fig:tf_hg0_thinfilm}
    } \hfill
    \subfloat[Pump mode: HG\(_1\)]{%
        \includegraphics[width=0.31\textwidth]{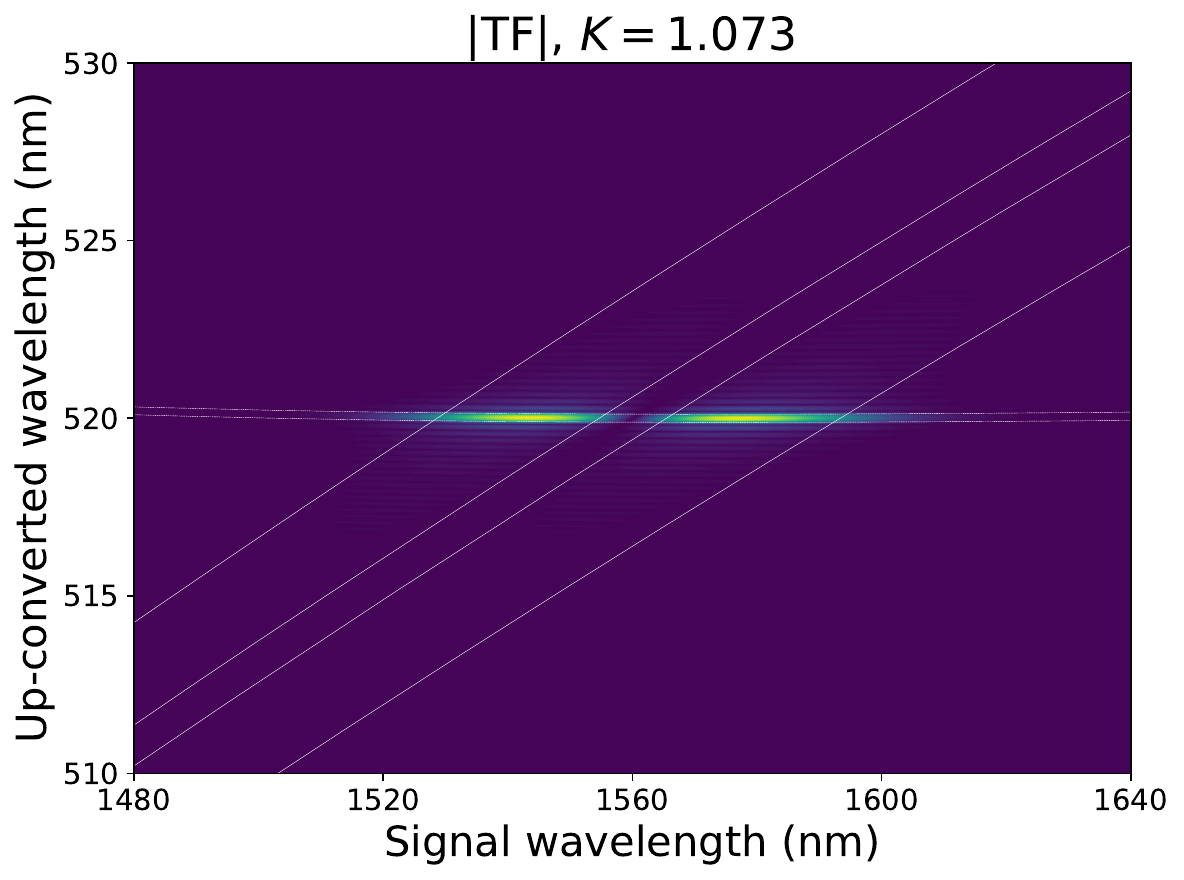}
        \label{fig:tf_hg1_thinfilm}
    } \hfill
    \subfloat[Pump mode: HG\(_2\)]{%
        \includegraphics[width=0.31\textwidth]{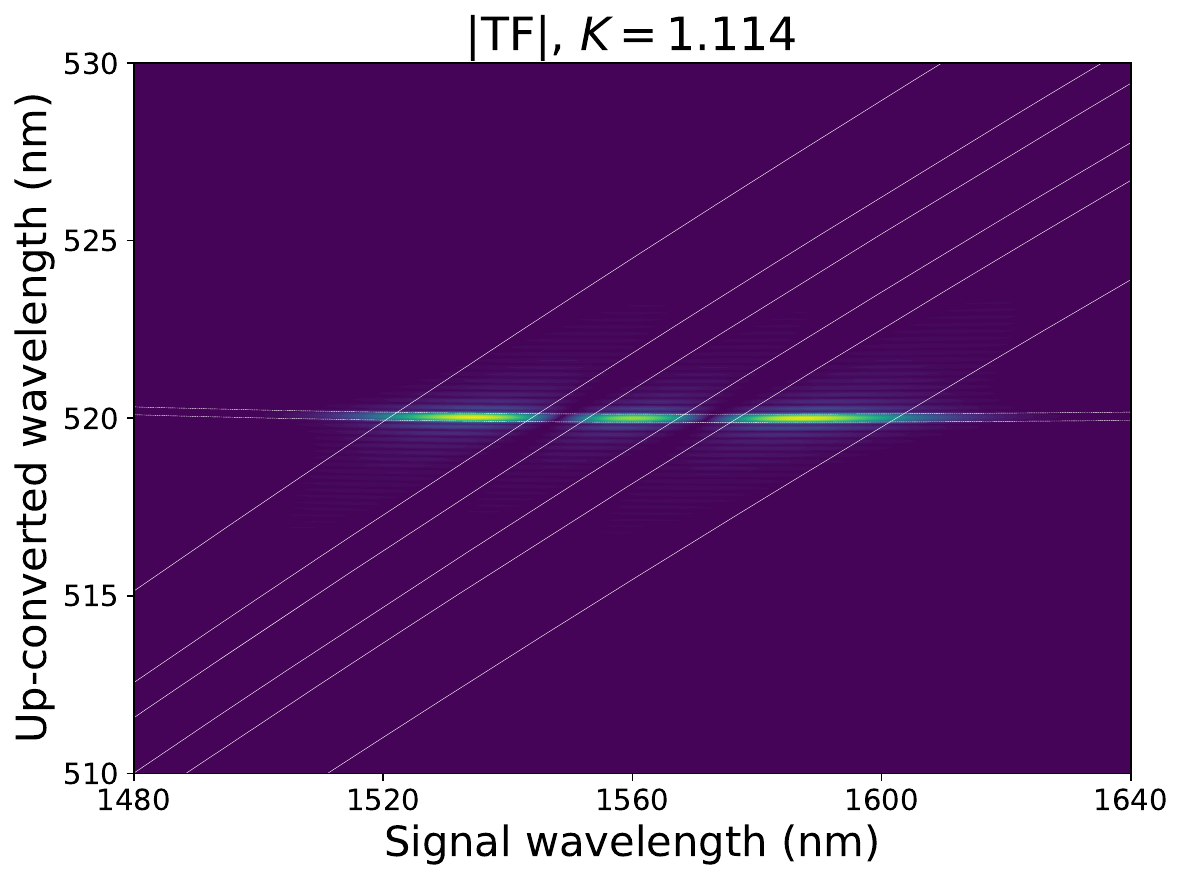}
        \label{fig:tf_hg2_thinfilm}
    }

    \caption{Transfer functions (TFs) for HG\(_0\), HG\(_1\), and HG\(_2\), corresponding to the parameters in Table~\ref{tab:SPS_thinfilm_subtraction}. The decomposed signal and up-converted modes are shown in Fig. \ref{fig:modes_SPS_thinfilm_group}.}
    \label{fig:tf_thinfilm_group}
\end{figure}

\begin{figure}[!htbp]
    \centering
    \subfloat[Obtained signal mode with pump: HG\(_0\)]{%
        \includegraphics[width=0.48\textwidth]{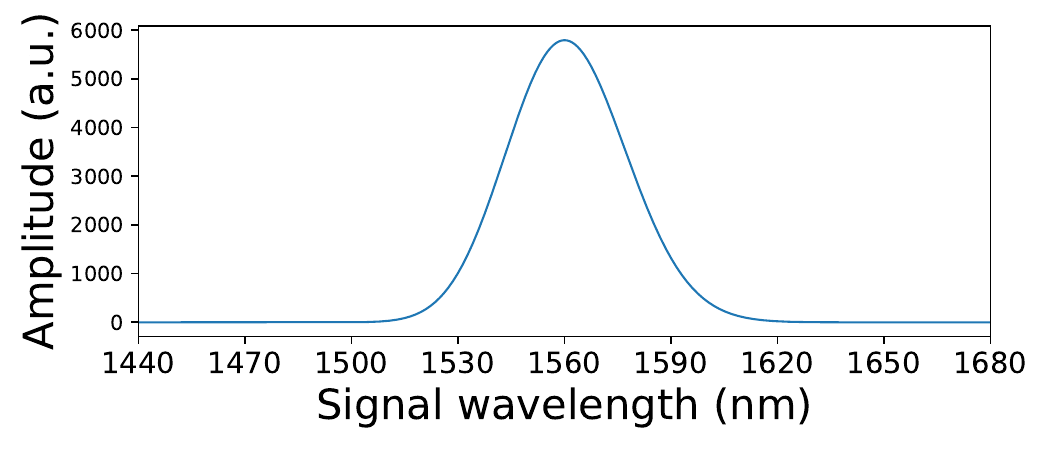}
        \label{fig:modes_sps_hg0_thinfilm}
    } \hfill
    \subfloat[Obtained signal mode with pump: HG\(_1\)]{%
        \includegraphics[width=0.48\textwidth]{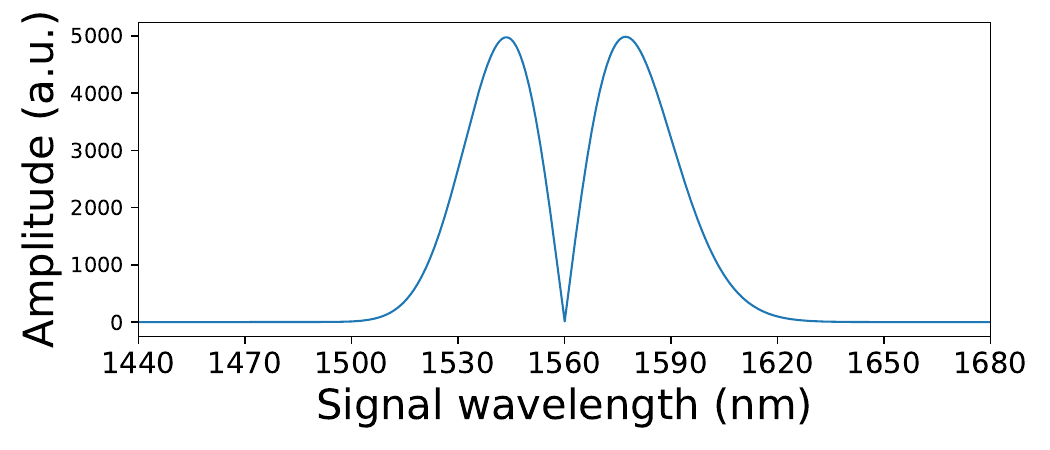}
        \label{fig:modes_sps_hg1_thinfilm}
    } \vspace{0.5cm} 
    \subfloat[Obtained signal mode with pump: HG\(_2\)]{%
        \includegraphics[width=0.48\textwidth]{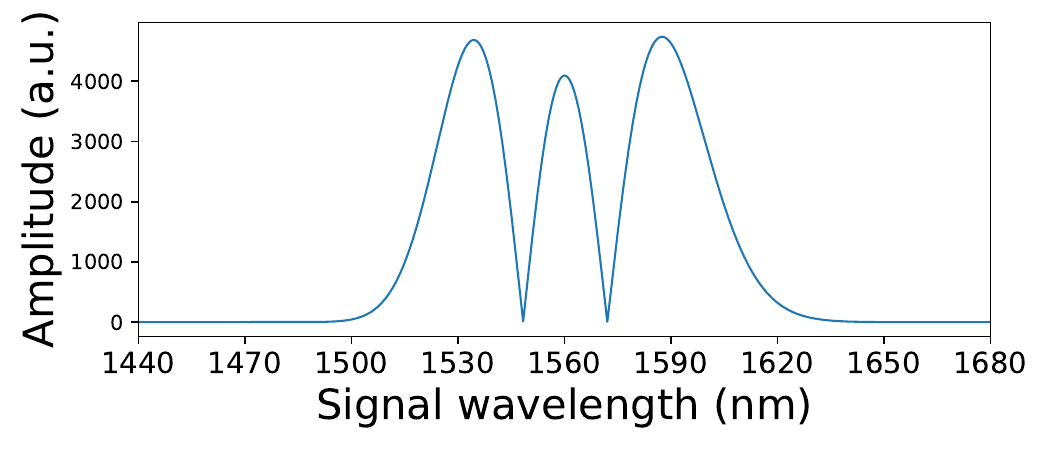}
        \label{fig:modes_sps_hg2_thinfilm}
    } \hfill
    \subfloat[Obtained gate mode for HG\(_0\)-pump. Representative for all gate-modes.]{%
        \includegraphics[width=0.48\textwidth]{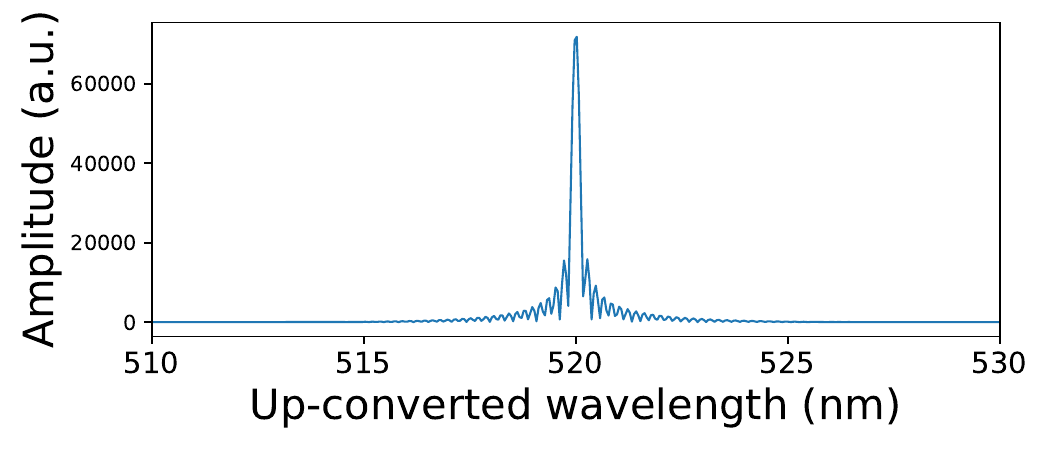}
        \label{fig:modes_sps_idler_thinfilm}
    }
    \caption{Signal and up-converted modes corresponding to the TFs shown in Fig.~\ref{fig:tf_thinfilm_group}, obtained using the parameters listed in Table~\ref{tab:SPS_thinfilm_subtraction}.}
    \label{fig:modes_SPS_thinfilm_group}
\end{figure}

Following the same approach, we now present the results for single-photon subtraction. The optimized waveguide and pump parameters yield the desired transfer functions (TFs) for HG pump modes of orders 0, 1, and 2, as shown in Fig.~\ref{fig:tf_thinfilm_group}. The corresponding signal and up-converted modes are presented in Fig.~\ref{fig:modes_SPS_thinfilm_group}. Since the up-converted mode remains nearly unchanged across all incoming HG orders, we present only the zero-order signal mode as a representative example.

\newpage
\section{Conclusion}
In this work, we investigated mode-selective single-photon addition and subtraction using two complementary waveguide modeling approaches. Starting from an analytically derivable metallic waveguide approximation—representative of indiffused waveguides on the micrometer scale—we extended our optimization scheme to a self-developed model of subwavelength-scale thin-film lithium niobate (TFLN) waveguides. The strong potential for dispersion engineering through geometric design in this platform enabled us to directly propose configurations suitable for implementing SPA and SPS processes.

By optimizing the waveguide–pump parameters in each case, we identified parameter sets yielding the desired Joint Spectral Amplitudes and Transfer Functions, thus enabling mode-selective single-photon addition and subtraction and facilitating the generation of non-Gaussian quantum states. For both processes, we found configurations that can serve as standalone experimental proposals. In the case of the fundamental Hermite-Gauss pump mode (HG\(_0\)), Schmidt numbers below \( K = 1.07 \) were found for SPA using both the metallic and TFLN waveguide models. For SPS, the optimization yielded Schmidt numbers of approximately \( K = 1.04 \) for the metallic model and \( K = 1.03 \) for TFLN. In all cases, higher-order pump modes (HG\(_1\), HG\(_2\)) led to similarly satisfactory results, confirming the robustness of the identified solutions with respect to mode-selectivity.

These findings demonstrate that, beyond the well-established metallic waveguide framework, the TFLN platform also constitutes a highly promising source for non-Gaussian state generation. Both platforms enable the selective generation of non-Gaussian states in a specific mode within a multimode input, using a single, static device configuration. The implementation is based on heralding via photon detection, which, to date, is the most efficient procedure for generating complex states in continuous variables \cite{Larsen2025,Winnel24,Konno24, Anteneh24,Simon24,Sychev17, Etesse15,Gerrits10,Takahashi08, Ourjoumtsev07,Neergaard-Nielsen06}. Here, we control the spectro-temporal mode where such an operation is applied \cite{Averchenko14,Averchenko16, Ra17,Roeland22,Biagi22}, which is particularly relevant since full scalability requires involving a large number of spectro-temporal modes of the field \cite{Bourassa2021}. Moreover, this specific implementation preserves coherence among multiple modes and enables the exploitation of the interplay between non-Gaussianity and quantum entanglement \cite{Ra20,Lopetegui25}.

\subsection*{Funding}
This   work   was   supported   by   the   European   Research Council under the Consolidator Grant COQCOoN (Grant No.  820079), by  Agence Nationale de la Recherche (OQuLus, ANR-22-PETQ-0013) and by Deutsche Forschungsgemeinschaft (DFG, German Research Foundation) – SFB-Geschäftszeichen TRR142/3-2022 – Projektnummer 231447078. P. N. was supported by QuantEdu-France (ANR-22-CMAS-0001) in the framework of France 2030.
\subsection*{Disclosure}
The authors declare no conflicts of interest.
\subsection*{Data availabilty}
Data and code underlying the results presented in this work are not publicly available at this time but may be obtained from the authors upon reasonable request.

\bibliography{main_arxiv}

\section*{APPENDIX}
\section*{Analytics of multimode single-photon subtraction and addition}
\label{sec:math_states_purity}
Following the approaches developed in the works of V. Averchenko \cite{Averchenko14, Averchenko16} and G. Roeland \cite{Roeland22}, we present a unified description of the mathematical frameworks to describe the processes of single-photon subtraction (SPS) and single-photon addition (SPA). This formalism provides a consistent basis for analyzing these quantum operations in a beam-splitter-based scheme as well as a PDC scheme.

\subsection*{Modeling Single-Photon Subtraction (SPS) as a Weak Beam Splitter}
To begin, we consider a pure, single mode, SPS process, to then treat the more general case. Such a process can be modelled by an input state undergoing a weak beam-splitter operation. This can be described by the evolution $\hat{U}:$
\[
\hat{U} = \text{exp}(i\theta(\hat{A}\hat{B}^\dagger-\hat{A}^\dagger \hat{B})) \approx 1 + i \theta (\hat{B}^\dagger \hat{A} - \hat{B} \hat{A}^\dagger),
\]
where a weak coupling regime is assumed for the right hand side approximation, implying that only single photon exchanges between signal and idler take place. 
The initial state in our SPS-scheme is given by
\(
\hat{\rho}_\text{in} = \hat{\rho}_s \otimes |0\rangle_B \langle 0|.
\), where \(\hat{\rho}_s\) is the state of the input signal, while the idler mode is on the vacuum.  To perform the single photon subtraction from the signal mode, post-selection on a single photon going to the idler is required. The detection event is described by the POVM 
\(
\Pi_B = |1\rangle \langle 1|_B,
\)
where \(|1\rangle _B,
\) is the single photon Fock state in the idler mode. Ideally, a photon number resolving detector should be used, but in practice click detectors suffice in the weak coupling regime that we assume. The appropriate POVM for such a detector is \(
\Pi_B^{\text{click}} = \mathrm 1 - |0\rangle \langle 0|_B,
\) which, given that only single photon events play a significant role, effectively acts as \(
\Pi_B^{\text{click}} = \Pi_B.
\)\\
The output state after a successful heralding of in the up-converted mode is given by 
\[
\hat\rho_{out} = \frac{1}{P} \text{Tr}_B \left( \underset{\hat{\rho}_\text{out}^\prime}{\underbrace{\hat{U} \hat{\rho}_\text{in} \hat{U}^\dagger}} \Pi_B \right),
\]
where $P$ describes the probability of successfully measuring a heralding photon, and thus subtracting it from the signal.
The state before heralding is given by $\hat{\rho}^\prime_{out}$:
\[
\hat{\rho}^{\prime}_\text{out} =
\hat{\rho}_s \otimes |0\rangle_B \langle 0|
+ i \theta ( \hat{A} \hat{\rho}_s \otimes |1\rangle_B \langle 0|)
- i \theta (\hat{\rho}_s \hat{A}^\dagger \otimes |0\rangle_B \langle 1|)
+ \theta^2 \hat{A} \hat{\rho}_s \hat{A}^\dagger \otimes |1\rangle_B \langle 1|.
\]
Therefore, for a single-photon subtracted state we obtain:
\[
\hat{\rho}_{out}=\frac{1}{P}\text{Tr}_B(\hat{\rho}_\text{out} \Pi_B) = \frac{\theta^2}{P} \hat{A} \hat{\rho}_s \hat{A}^\dagger
\]
with $P=\theta^2\text{Tr}[\hat{A}^\dagger\hat{A}\hat{\rho}_{s}]=\theta^2 \bar n_{s}$, the probability of the process to detect a photon in the heralding path, and $\bar n_{s}$ the average photon number of the input state. Notice that there are two ways to boost the probability of success: increasing the average photon number of the input (equivalently, the amount of squeezing in the target mode), or having a stronger coupling (higher beam splitter transmissivity, or stronger pump and non-linearity in the case of up convertion process). Nevertheless, this cannot be done arbitrarily as otherwise higher photon contributions may appear. The non-linear term is effectively not in the equation for the output state as it appears as a multiplicative factor, meaning we can write the state as $\hat{\rho}_{out}=\frac{1}{\mathcal N_A} \hat{A} \hat{\rho}_s \hat{A}^\dagger$, where $\mathcal N_A=\bar n_s$.

\subsection*{Purity of the SPS-process}
The experimental implementation of single photon subtraction is, on the other hand, a non-pure process. While the beam splitter approach provides an appealing way of deriving the effect of single photon subtraction, the non-linear process through which we implement this process is more subtle. Given that the photon detection scheme is not mode selective, and several signal modes are coupled through the non-linear process to the idler, a successful heralding event is effectively a statistical mixture of subtracted states. In what follows we characterize the purity of these states. \\ 
The purity $\mu^-$ of a state is defined as $\mu^- = \text{Tr}[\hat{\rho}_{out}^2]$. 
Since in this work we assume pure multimode states as an input, we can characterize the purity of the output states by simultaneously describing the purity in mode-selectivity of the nonlinear process per se.  \\ We start by considering an input multimode state $\hat{\rho_{in}} =\ket{\phi}\bra{\phi} =\bigotimes_{i=1}^n \ket{\phi_i}\bra{\phi_i}$. In all simulated scenarios the Schmidt decomposition of the JSAs/TFs results in several non-zero eigenvalues, which means that the process not only involves our target mode, but also other modes orhthogonal to it. The Hamiltonian of the non-linear process can be written as 
\[
\hat H= \int \int d \omega_s d\omega_{up}\hat a(\omega_s)\hat b^\dagger(\omega_{up})+h.c.
\]
After performing a Singular value decomposition on the transfer function, 
\[T(\omega_s,\omega_{up})=\sum_l \sqrt{\lambda_l}m_l(\omega_s)n_l(\omega_{up}),\]
the process can be written as 
\[
\hat H= \sum_l \sqrt{\lambda_l} (\hat A_l \hat B^\dagger_l+h.c.),
\]
where $\hat A_l=\int d\omega_s m_l(\omega_s)\hat a(\omega_s) $, and  $\hat B_l=\int d\omega_{up} n_l(\omega_{up})\hat b(\omega_{up})$, are the annihilation operators of the eigenmodes of the process $\{m_l(\omega_s)\}, \{n_l(\omega_{up})\}$. Following a calculation similar to the one performed in the single mode case, we obtain the output state
\[
\hat{\rho}_{out} = \frac{1}{P} \sum_l \lambda_l A_l \hat{\rho_{s}} A_l^\dagger,
\]
where $P=\sum_l \lambda_l \bar n_l$, is the success probability of the photon subtraction. Notice that, as discussed on the introduction of this section, the output state is given by a mixture of photon subtractions in different modes, due to the unavoidable multimode nature of the process and the non mode-selective photon counting. Nevertheless, for most cases of interest, we obtain a value of $\lambda_1$ significantly higher than the rest, meaning that the contribution from subtractions in undesired modes is minimal. We recall here that we consider the input state to be a pure state $\hat \rho_s=|\phi_s\rangle\langle\phi_s|$. This allows us to use, as a figure of merit for the mode selectivity of the process, the purity of the output state, which is given by   
\[
\text{Tr}[\hat{\rho}_{out}^2] = \frac{1}{P^2}  \left( \sum_l \lambda_l^2\bar n_l^2 +\sum_{l\neq k}\lambda_l \lambda_k \left|\langle\phi_s|\hat A_l^\dagger\hat A_k|\phi_s\rangle\right|^2 \right).
\]

Defining the normalized coefficients:
\[
\tilde{\lambda}_i = \frac{\lambda_i}{P},
\]
we obtain:
\[
\mu^-=\text{Tr}(\hat{\rho}_{out}^2) =  \sum_{l,k}\tilde\lambda_l \tilde\lambda_k \left|\langle\phi_s|\hat A_l^\dagger\hat A_k|\phi_s\rangle\right|^2 .
\]

To figure out a neat form for this evaluation we recall that the input state, $|\phi_s\rangle$, is given by 
\[
|\phi_s\rangle=\bigotimes|\phi_i\rangle,
\]
where 
\[|\phi_i\rangle=\frac{1}{\sqrt{\cosh r_i}}\sum_n \frac{\sqrt{(2n)!}}{2^n n!}\left(e^{-i \xi_i}\tanh(r_i)\right)^n |2n\rangle,\]
where $r_i,\xi_i$ are the squeezing level and phase respectively. The basis in which the input state is written as a tensor product $\{\phi_i(\omega_s)\}$, is not necessarily the same as the SVD basis of the photon subtraction process, $\{m_l(\omega_s\}$. Because of this, contributions from crossed terms on the purity cannot be dismissed. The eigenmodes of the photon subtraction process are related to the squeezed modes by $O_{ij}=\int d\omega_s m_i(\omega_s)\phi_j(\omega_s)$, such that $\hat A_l=\sum_n O_{ln}\hat a_n$, where $\hat a_n$ is the annihilation operator corresponding to the $n^{th}$ squeezed state. As a consequence, 
\[
\mu^-= \sum_{l,k}\tilde\lambda_l \tilde\lambda_k \left|\sum_{m,n} O_{lm} O_{kn} \langle \phi_s|\hat a^\dagger_m \hat a_n|\phi_s|\rangle\right)^2, 
\]
which, taking into account taht for vacuum squeezed input states $\langle\phi_i|\hat a_i |\phi_i\rangle=0$, implies that  
\[
\mu^-= \sum_{l,k}\tilde\lambda_l \tilde\lambda_k \left(\sum_m O_{lm} O_{km} \bar n_m\right)^2 .
\]
 On the other hand, if the states are bright squeezed state, i.e. if they have non zero mean field, then the overlap $\langle\phi_i|\hat a_i |\phi_i\rangle\neq 0$ and more contributions have to be considered. \\

\subsection*{Modeling Single-Photon Addition (SPA) as a PDC-process}
While the SPS process is modeled as a weak beam-splitter interaction, the SPA process can be described by a parametric down-conversion (PDC) process, as sketched in \cref{fig:experimental_setups}a. The Hamiltonian of the process is given by
\[
\hat{H} = \int \int d\omega_s d\omega_i \, J(\omega_s, \omega_i) \hat{a}^\dagger(\omega_s) \hat{b}^\dagger(\omega_i) + \text{h.c.}
\]
Doing SVD on the JSA $J(\omega_s,\omega_i)=\sum_l \sqrt{\lambda_l} h_l(\omega_s)g_l(\omega_i)$, we obtain the eigenmode decomposition of the process
\[
\hat H= \sum_l \sqrt{\lambda_l} \hat A^\dagger_l\hat B^\dagger_l + h.c.,
\]
where $\hat A^\dagger_l=\int d \omega_s h_l(\omega_s) \hat a(\omega_s)$, and $\hat B^\dagger_l=\int d \omega_i g_l(\omega_i) \hat b(\omega_i)$, are the creation operator of the eigenmodes of the process. 
Under similar assumptions, and following a similar procedure as for the single photon subtraction case, we can show that the output state is given by
\[
\hat{\rho}_\text{out}^+ = \frac{1}{P}\sum_l \lambda_l  \hat A^\dagger_l\hat \rho \hat A_l, 
\]
where $P$ is the probability of detecting a photon in the idler mode:
\[
P = \text{Tr}(\sum_l \lambda_l \hat A^\dagger_l\hat \rho \hat A_l)=\sum_l \lambda_l \left(1+\langle\phi_s|\hat A^\dagger_l\hat A_l|\phi_s\rangle\right),
\]
which, considering the expansion of the eigenmodes of the process in terms of the signal (Hermite Gauss) modes $\hat A_l=\sum_m O_{lm} \hat a_m$, and the fact that each of the signal modes is in a squeezed vacuum state, leads to 
\[
P=\sum_l \lambda_l + \sum_{l,m} \lambda_l O_{lm}^2\bar n_m.
\]
where \(\bar{n}_m\) denotes the mean photon number in the $m^{th}$ signal (Hermite Gauss) mode.\\\\
The purity of the outcoming state $\mu^+=\text{Tr}[\hat{\rho}_{out}^2]$, is thus given by:
\[
\mu^+ = \sum_{l,k}\tilde \lambda_l \tilde\lambda_k \left | \sum_m O_{lm}O_{km}(1+\bar n_m)\right|^2,
\]
where 
\(
\tilde{\lambda}_i = \frac{\lambda_i}{P},
\) are normalized coefficients.

\subsection*{Purity Evaluation of the Obtained Processes}

The previously derived expressions allow us to directly quantify the intrinsic purity of the obtained processes. One should keep in mind, as discussed in Sec.~2.5, that additional spectral filtering on the idler/up-converted field can increase the purity further towards unity. 
As highlighted in Sec.~2.7, the spectral width of the pump was chosen to strike a "sweet spot" between spectral shapability and a process Schmidt number of $K\approx1$. For us, a sufficient configuration was found to correspond to a pump width of 7~nm for the metallic waveguide approximation and 10~nm for the TFLN model in the SPS process. For the SPA process, 7~nm was chosen for the metallic waveguide approximation and 6~nm for the TFLN model.\\

\begin{table}[H]
\centering

\begin{minipage}[t]{0.47\textwidth}
\centering
\begin{tabular}{|l|l|l|}
\hline
       & $\mu^-_{\text{met}}$ & $\mu^-_{\text{TFLN}}$ \\
\hline
HG$_0$ & 0.969                & 0.976                 \\
HG$_1$ & 0.942                & 0.941                 \\
HG$_2$ & 0.919                & 0.909                 \\
\hline
\end{tabular}
\captionof{table}{SPS process}
\end{minipage}
\hfill
\begin{minipage}[t]{0.47\textwidth}
\centering
\begin{tabular}{|l|l|l|}
\hline
       & $\mu^+_{\text{met}}$ & $\mu^+_{\text{TFLN}}$ \\
\hline
HG$_0$ & 0.952                & 0.954                 \\
HG$_1$ & 0.884                & 0.895                 \\
HG$_2$ & 0.826                & 0.822                 \\
\hline
\end{tabular}
\captionof{table}{SPA process}
\end{minipage}

\vspace{1ex}

\caption{Process purities for the first three Hermite-Gauss pump/gate modes, containing the metallic waveguide approximation and the TFLN model for (a) the SPS process and (b) the SPA process.}
\label{tab:purity_results_all}
\end{table}

\section*{METHODS}
\subsection*{Functionality of the Genetic Algorithm}
\label{sec:methods_algorithm}

We implement a genetic algorithm, which optimizes a problem by iteratively improving a population of candidate solutions through the operations of \textit{selection}, \textit{crossover}, and \textit{mutation}. Each solution, referred to as an individual, is represented by a vector \( x \in \mathbb{R}^L \), where \( L \) denotes the number of parameters in the genome. Since the objective is to find a set of parameters that enable mode-selective single-photon addition/subtraction, each vector \( x \) encodes a specific configuration of tunable waveguide and pump properties.

\subsubsection*{\textbf{Initialization of the Population}}  
A population \( \mathcal{P} = \{x^{(i)}\}_{i=1}^{N} \) of \( N \) individuals is initialized. Each individual  
\[
x^{(i)} = (x^{(i)}_1, x^{(i)}_2, \dots, x^{(i)}_L) \in \mathbb{R}^L
\]
represents a candidate configuration, such as geometrical waveguide parameters and the spectral pump width. The components \( x^{(i)}_j \) are drawn independently from normal distributions \( \mathcal{N}(\mu_j, \sigma_j) \), subject to a hard constraint \( x^{(i)}_j \geq x_{j,\text{min}}\), ensuring manufacturing feasibility:
\[
x^{(i)}_j \sim \mathcal{N}(\mu_j, \sigma_j), \quad \text{with} \quad x^{(i)}_j \geq  x_{j,\text{min}}.
\]
\subsubsection*{\textbf{Parent Selection}}  
The goal of the optimization process is to iteratively move the population toward individuals with smaller fitness values, i.e., those closer to the target configuration. The fitness value, as described below, quantifies how close an individual is to the target configuration.

To this end, two parent individuals \( x^{(p1)}, x^{(p2)} \in \mathcal{P} \) are selected based on their fitness values. The selection is carried out via the \emph{tournament selection} method. In each selection round, a random subset \( T \subset \mathcal{P} \) of the current population is drawn, where \( |T| = N/2 \). The individual within the subset that exhibits the best fitness value \( f(x) \) is selected as a parent:
\[
x_{\text{best}} = \arg\min_{x \in T} f(x).
\]
This process is repeated independently to obtain both parents. The selection mechanism thus introduces a degree of stochasticity while preserving selective pressure toward fitter individuals.

\subsubsection*{\textbf{Crossover}}  
To generate new individuals, we apply a linear crossover between two selected parents. Given two parent vectors \( x^{(p1)} \) and \( x^{(p2)} \in \mathbb{R}^L \), two offspring \( x^{(c1)} \) and \( x^{(c2)} \) are created by forming weighted linear combinations of the parents. Specifically, each component \( i \in \{1, \dots, L\} \) of the offspring is computed as
\[
x^{(c1)}_i = \beta_1 \cdot x^{(p1)}_i + (1 - \beta_1) \cdot x^{(p2)}_i,
\]
\[
x^{(c2)}_i = \beta_2 \cdot x^{(p1)}_i + (1 - \beta_2) \cdot x^{(p2)}_i,
\]
where the coefficients \( \beta_1, \beta_2 \) are drawn independently and uniformly  from the interval $\left[-d,\; d+1\right]$, where $d$ is a hyperparameter, that we set to, for example, $d=0.5$. This mechanism allows both interpolation ($\beta_i\in\left[0,1\right]$) and extrapolation ($\beta_i<0$ or $\beta_i>1$) around the parent vectors, depending on the values of \( \beta_k \).

This method promotes genetic diversity by enabling exploration of a broader parameter space. To ensure physical validity, the procedure is repeated until both offspring satisfy the constraint \( x^{(c)}_i \geq \text{min}_i \) for all components \( i \in \{1, \dots, L\} \).

\subsubsection*{\textbf{Mutation}}  
To maintain genetic diversity and avoid convergence to local minima, random mutations are applied to individual genomes. With a fixed probability defined by the mutation rate, a given individual is selected for mutation. In that case, each component of the genome is perturbed by adding Gaussian noise with zero mean and a standard deviation matching the corresponding initialization value. Specifically, for each gene \( i \in \{1, \dots, L\} \), the mutated value is sampled as
\[
x^{\text{(mut)}}_i = x_i + \mathcal{N}(0, \sigma_i),
\]
with the additional constraint \( x^{\text{(mut)}}_i \geq \text{min}_i \). If any component fails to meet this constraint, the sampling is repeated until a valid configuration is obtained.
\subsubsection*{\textbf{Fitness Evaluation and Simulation}}

Each individual \( x^{(i)} \in \mathcal{P} \) is evaluated by simulating the nonlinear process defined by its parameters. These parameters are passed to the simulation to compute the resulting joint spectral amplitude (JSA) or transfer function (TF). From there, a fitness value is derived, quantifying how well the configuration meets the desired physical behavior — such as process purity.
The quality of a given configuration is quantified using the Schmidt number \( K \), which serves as an indicator of mode selectivity. The ideal value corresponds to \( K_s = 1 \), representing perfect single-mode behavior. The fitness value is defined as the absolute deviation from this ideal:
\[
f(x^{(i)}) = |K_s - K(x^{(i)})|.
\]

To ensure that the solution performs robustly across multiple relevant spectral profiles, the simulation was also carried out using the first three Hermite-Gauss pump modes as input. The fitness score in this case is computed as the average deviation across all test cases:
\[
f(x^{(i)}) = \frac{1}{3} \sum_{k=1}^{3} |K_s - K^{(k)}(x^{(i)})|,
\]
where \( K^{(k)}(x^{(i)}) \) denotes the Schmidt number obtained when using the \( k \)-th Hermite-Gauss pump mode. It should be noted that both approaches—evaluating a single mode or averaging over multiple modes—led qualitatively to similar results, as the orientation of the phase-matching function was found to be the primary limiting factor.

\subsubsection*{\textbf{Iterative Population Update}}  
The evolutionary process is carried out over \( G \) generations. In each generation, a subset of individuals is evaluated based on their fitness, and parent individuals are selected via tournament selection. Crossover and mutation are then applied to generate new offspring. The resulting population fully replaces the previous one, forming the basis for the next generation. Over successive generations, the optimization gradually converges toward configurations exhibiting strong mode selectivity, as indicated by consistently low fitness values in the best-performing individuals.

\end{document}